\documentclass{aa}  

\usepackage{graphicx}
\usepackage{txfonts}
\usepackage[colorlinks=True,pdfborder={0 0 0}, linkcolor=blue,citecolor=blue]{hyperref}

\def\tvi(#1,#2){\vrule height #1pt depth #2pt width 0pt}

\def\v2{v_{\rm sh}}

\def\eg{e.g. }

\def\M{{\cal M}}

\begin{document}

   \title{Analytic insight into the physics of SASI}

   \subtitle{II. Spiral instability of the prograde mode in a rotating stellar core}

   \author{T. Foglizzo
          }

   \institute{Universit\'e Paris-Saclay, Universit\'e Paris Cit\'e, CEA, CNRS, AIM, 91191, Gif-sur-Yvette, France\\
              \email{foglizzo@cea.fr}
             }

   \date{Received 4 June 2026; accepted 13 July 2026}
 
  \abstract
  {During the core-collapse of a rotating massive star, the standing accretion shock instability (SASI) favours the development of non-axisymmetric motions which can imprint specific frequency signatures on the neutrino and gravitational wave signals. }
   {This study establishes analytical approximations for the eigenfrequencies of the dominant SASI modes. It also explains the physical mechanism responsible for the further destabilization of prograde SASI modes by differential rotation. }
  {A perturbative analysis is used to calculate the eigenfrequencies of a stalled accretion shock in spherical geometry, taking into account the rotation of the collapsing stellar core. The formulation of the perturbative equations as a self-forced oscillator is extended to include the effect of differential rotation and interpret the results physically. }
   {The oscillation frequency of the dominant mode weakly depends on the detailed formulation of neutrino emission if the shock radius exceeds $\sim 1.5$ times the radius $r_\nabla$ of maximum deceleration. Analytical expressions are obtained for the one and two-armed spiral modes with a $10\%$ accuracy in this regime.
The effect of differential rotation on SASI is explained by the role of phase mixing between the advective forcing and the acoustic structure.
The radial wavelength of vorticity perturbations associated with the prograde mode is increased by differential rotation, leading to a better phase match with the large radial scale of the acoustic structure. Even when rotation is too modest to involve a corotation radius, its adverse effect on phase mixing can be significant at small radius due to the steep inward increase of the rotation frequency $\propto 1/r^2$ in the region of stationary accretion. In the regime of faster rotation involving a corotation radius, the stationary phase approximation sheds light on the dominant advective-acoustic coupling, located between the corotation zone and the shock. 
}
   {}

   \keywords{hydrodynamics -- instabilities -- shock waves -- stars: neutron -- supernovae: general -- neutrinos
               }

   \maketitle

\section{Introduction}

The explosive death of massive stars is sensitive to the development of hydrodynamical instabilities which break the spherical symmetry of the stellar core, affect the efficiency of neutrino energy absorption and generate turbulence \citep{Muller2020,Janka2016,Burrows2021}. A spherically symmetric scenario based on radial motions seems possible only for the lightest progenitors \citep{Kitaura2006,Stockinger2020}.  Asymmetric motions contribute to the kick and spin of the neutron star \citep{Muller_Tauris2019, Janka2017}, the emission of gravitational waves \citep{Kotake2017} and a modulation of neutrino emission \citep{Muller2019a,Tamborra2019}. Ultimately, the information encoded in the gravitational waves and neutrino signals can be used to recover the properties of the dying star and its explosion mechanism \citep{Powell2022}. 
The computational cost of 3D numerical simulations precludes a systematic coverage of the large parameter space describing the initial conditions of the stellar core-collapse. Understanding the underlying mechanism of instabilities is essential to extrapolate the results of sparse numerical simulations, evaluate the impact of additional physical ingredients and design effective prescriptions for parametric studies  \citep{Muller2019b}.\\
Among the hydrodynamical instabilities at work during the phase of stalled accretion shock, the Standing Accretion Shock Instability \citep{Blondin2003} is able to introduce coherent transverse motions with a large angular scale, growing over a timescale related to the advection time from the shock to the neutron star surface.
The mechanism of SASI has been described as an advective-acoustic cycle between the shock and the vicinity of the proto-neutron star \citep{Foglizzo2007, Fernandez2009a, Guilet2012}. 

The rotation of the stellar core was considered in the first 3D simulations of SASI which revealed that its prograde spiral mode is favoured \citep{Blondin_Mezzacappa2007}.
A perturbative analysis by \cite{Yamasaki2008} (hereafter YF08) demonstrated that the growth rate of the prograde mode increases approximately linearly with the specific angular momentum of the collapsing core even in the regime of moderate rotation where centrifugal effects are quadratic and negligible. This linear effect was analytically associated with differential rotation, but its physical mechanism remains unexplained. {\it How can the growth rate of SASI be significantly enhanced by rotation for rotation rates slower than the SASI frequency?}

 The significantly different advection timescales in spherical and cylindrical geometries led \cite{Walk2023} (hereafter WFT23) to challenge our understanding of the SASI mechanism 
when rotation is included. According to their perturbative study, the oscillation frequency associated with SASI becomes less sensitive to the advection time from the shock down to the vicinity of the neutron star surface when the collapsing core is rotating. 

Even though it is not fully understood, the destabilizing effect of rotation on the prograde mode of SASI was confirmed by the 3D numerical simulations of \cite{Iwakami2009a,Iwakami2009b} who included a realistic equation of state and neutrino heating in a  spherical geometry. The perturbative results of YF08 have been directly tested and confirmed by \cite{Blondin2017} using 2D numerical simulations in cylindrical and spherical-equatorial geometries as well as 3D simulations in spherical geometry. A limitation of perturbative studies of SASI including rotation is the approximate character of the equatorial approximations which neglect the poloidal structure. The cylindrical approximation used by Y08 and the spherical-equatorial approximation used by \cite{Walk2023} and \cite{Buellet2023} capture the effect of the azimuthal wavenumber $m$, but do not take into account the poloidal wavenumber $\ell$ of spherical harmonics. This difficulty limits the accuracy of perturbative results and their ability to interpret the SASI frequencies in numerical simulations of the collapse of a rotating progenitor.  Analytical approximations of the dominant SASI frequency have been proposed in non-rotating progenitors \citep{Muller_Janka2014,Torres2021,Foglizzo2024} but none so far have included the effect of rotation.

The main progress in \cite{Foglizzo2024} (herafter Paper~I) was to note that the SASI frequency of the fundamental $\ell=1$ mode, in addition to being sensitive to the shock radius $r_{\rm sh}$, is sensitive to the radius $r_\nabla$ of maximum deceleration rather than the radius of the proto-neutron star. The physical mechanism of SASI was further clarified by describing the post-shock acoustic region as a self-forced oscillator.

The present paper extends this approach to include the effect of rotation.
The difficulty of solving the poloidal structure in a rotating progenitor is overcome in Sect.~2.2 by using a conservation property of the radial vorticity, and solved numerically in Sect.~2.3. The analytical approximation of SASI frequency obtained in Paper~I for the fundamental mode $\ell=1$ is extended to take into account rotation and also extended to the mode $\ell=2$ relevant to gravitational waves. This analytical approximation is compared to state-of-the-art formulae without rotation in Sect.~2.4.
The adiabatic approximation leading to an integral equation for the eigenfrequencies is established in Sect.~3 with rotation. The magnitude of non-adiabatic effects is evaluated by comparing them with the non-adiabatic results of Sect.~2. The consequences on our understanding of SASI are described in Sect.~3.4 for moderate rotation, and in Sect.~3.5 with a corotation radius. The main results are summarized in Sect.~4.1 and perspectives are discussed in Sect.~4.2. 

\section{Equatorial accretion with a cooling function and differential rotation}

This section establishes the equations describing the non-adiabatic stationary flow in the equatorial plane and its linear perturbations. The solutions are computed numerically and approximated analytically.

\subsection{Stationary flow}

As in WFT23 and Paper~I our simplified model describes the phase where the accretion shock stalls at the radius $r_{\rm sh}$, while neutrinos are emitted near the proto-neutron star radius $r_{\rm ns}$. 
Neutrino absorption is neglected and cooling by neutrino emission is idealized with a cooling function ${\cal L}$ depending on the local density $\rho$ and pressure $p$:
\begin{eqnarray}
{\cal L}\equiv A_{\rm cool}\rho^{\beta-\alpha} p^\alpha.\label{def_cool}
\end{eqnarray}
The collapsing stellar core immediately after bounce is modelled as a perfect gas with an adiabatic index $\gamma=4/3$, dominated by the degeneracy pressure of relativistic electrons. 
The dimensionless measure of the entropy $S$ is defined by:
$S\equiv (\log{p/\rho^\gamma})/(\gamma-1)$.
The gravitational potential $\Phi\equiv -GM/r$ is assumed to be dominated by the mass $M$ of the proto-neutron star in the Newtonian approximation. The self-gravity of the accreting gas and the increase of $M$ with time are neglected.\\ 
In order to study the effect of rotation we focus on the vicinity of the equatorial plane and consider a radially uniform distribution of specific angular momentum $L\equiv r v_\varphi $. The reference angular momentum $L_{\rm K}^r$ is defined as the Keplerian specific angular momentum at the radius $r$:
\begin{eqnarray}
L_{\rm K}^r\equiv (GMr)^{\frac{1}{2}}.
\end{eqnarray}
$L_{\rm K}^{\rm ns}$ is thus an upper bound for stationary accretion. The equatorial stationary flow is described by the mass conservation, the entropy equation, the radial and azimuthal components of the Euler equation:
\begin{eqnarray}
     \rho v_{r} &=& \left(\frac{r_{\rm sh}}{ r}\right)^2\rho_{\rm sh} v_{\rm sh}\ , \label{eq_mass_stat}\\ 
    \frac{\partial S}{\partial r} &=& \frac{{\cal L}}{ p v_r}\ , \label{eq_entropy_stat}\\
    \frac{\partial}{\partial r} \left( \frac{v_r^2}{ 2}  + \frac{c^2}{\gamma - 1} + \Phi+\frac{L^2}{ 2r^2} \right) &=& \frac{\cal L}{\rho v_r}\ ,\label{eq_euler_stat}\\
    \frac{\partial L}{\partial r}&=&0,\label{eq_eulerphi}
\end{eqnarray}
where $c\equiv (\gamma p/\rho)^{1/2}$ is the sound speed. The subscript "sh" in Eq.~(\ref{eq_mass_stat}) refers to quantities immediately below the shock, and the subscript "1" refers to pre-shock quantities.
The jump conditions at the shock follow from the conservation of mass flux and momentum flux. The energy lost across the shock through nuclear dissociation can be taken into account with the same parameter $\varepsilon\equiv {2\Delta e_{\rm disso}/v_1^2}$ as in Paper~I, but is neglected in the numerical calculations of this study.
Assuming that the incoming gas is cold with a negligible pre-compression, the centrifugal correction to the radial velocity $v_1$ depends quadratically on $L$ according to the Bernoulli equation:
\begin{eqnarray}
\eta^2&\equiv &\frac{v_1^2r_{\rm sh}}{ 2GM},\label{def_eta}\\
&\sim&1-\left(\frac{L}{ L_{\rm K}^{\rm ns}}\right)^2\frac{r_{\rm ns}}{ 2r_{\rm sh}}.\label{eqapeta2}
\end{eqnarray} 
The present study is focused on the regime $L/L_{\rm K}^{\rm ns}<0.5$ where centrifugal effects are minor. In this regime the rotation is considered "moderate" if the rotation period at the inner boundary is longer than the advection timescale $r_{\rm sh}/|v_{\rm sh}|$. The "corotation regime" refers to a rotation period shorter than the advection timescale. Defining the Mach number $\M\equiv -v_r/c$ as positive and associated with the radial component of velocity, we assume a strong adiabatic shock $\M_1\gg1$ in the numerical calculations of this section.

\subsection{Perturbed flow}

The perturbations of the stationary flow are characterized by spherical harmonics $Y_\ell^m$ with a poloidal number $\ell$, an azimuthal number $|m|\le \ell$,  and a complex eigenfrequency $\omega$. The linear perturbations are defined as $\delta h$ for the mass flux, $\delta S$ for the dimensionless entropy, and $r\delta v_\varphi$ for the specific angular momentum. The poloidal component of the vorticity equation (Eq.~(\ref{dwthetadr}) in Appendix~\ref{append_nonax}) involves a combination of entropy $\delta S$ and vorticity $\delta w_\theta$ which we name $\delta k_\theta$, related to the baroclinic production of vorticity:
\begin{eqnarray}
\delta h &\equiv& \frac{\delta v_r}{ v_r} + \frac{\delta \rho}{\rho} \ , \label{defdh0}\\
\delta S &\equiv& \frac{1}{\gamma-1} \frac{\delta c^2}{c^2} - \frac{\delta \rho}{\rho}\ , \label{defdS0}\\
\frac{\delta k_\theta}{ im}&\equiv&  \frac{c^2}{ \gamma}\delta S+\frac{rv_r}{ im}\delta w_\theta.\label{FSW}
\end{eqnarray}
By focusing on perturbations which are symmetric with respect to the equatorial plane, we show in Appendix~\ref{append_nonax} that velocity and density perturbations are related to the variables $\delta h,r\delta v_\varphi, \delta k_\theta,\delta S$ as follows:
\begin{eqnarray}
\frac {\delta v_r }{v_r }=\frac{1}{ 1-\M^2}\left(
 \delta h
- \frac{i\omega' }{ c^2} \frac{r\delta v_\varphi}{ im}
- \frac{\delta k_\theta}{ imc^2}
+ \delta S
 \right),\label{dvr}\\
\frac{\delta \rho}{\rho}=\frac{1}{ 1-\M^2}\left(
-\M^2\delta h
+\frac{i\omega' }{c^2} \frac{r\delta v_\varphi}{ im}
+  \frac{\delta k_\theta}{ imc^2}
-  \delta S \right),\label{drho}
\end{eqnarray}
where the Doppler shifted frequency $\omega'$ depends on the azimuthal wavenumber $m$ and the rotation frequency $L/r^2$. We introduce the same Lamb frequency $\omega_{\rm Lamb}$ as in spherical symmetry. It involves the poloidal wavenumber $\ell$ associated with the angular part of the Laplacian: 
\begin{eqnarray}
\omega'&\equiv& \omega -\frac{mL}{ r^2},\label{def_omprime}\\
\omega_{\rm Lamb}^2&\equiv& \ell(\ell+1)\frac{c^2-v_r^2}{r^2}.
\end{eqnarray} 
The differential system satisfied by $(\delta h,r\delta v_\varphi, \delta k_\theta,\delta S)$, derived in Appendix~\ref{append_nonax}, is expressed with $\omega'$ and $\omega_{\rm Lamb}$:
\begin{eqnarray}
\left(\frac{1-\M^2}{ v_r}\frac{\partial}{\partial r}+\frac{i\omega' }{c^2}\right)\delta h=
-\frac{\omega'^2-\omega_{\rm Lamb}^2}{ v_r^2c^2} \frac{r\delta v_\varphi}{im}\nonumber\\
+\frac{i\omega' }{v_r^2}\left(
\frac{\delta k_\theta}{ imc^2}
-\delta S
\right)
,\label{ddhdr}
\\
\left(\frac{1-\M^2}{ v_r}\frac{\partial}{\partial r}+\frac{i\omega' }{c^2}\right)\frac{r\delta v_\varphi}{ im}
=
\delta h
-\frac{\delta k_\theta}{ imv_r^2}
+\left(\gamma-1+\frac{1}{\M^2}\right)\frac{\delta S}{\gamma}
,\label{ddLdr}
\\
\left(\frac{\partial}{\partial r}-\frac{i\omega'}{v_r}\right)
\frac{\delta k_\theta}{ im}
=
\delta\left(\frac{{\cal L}}{\rho v_r}\right),\label{ddkdr}
\\
\left(\frac{\partial}{\partial r}-\frac{i\omega'}{v_r}\right)\delta S
=
\delta\left(\frac{{\cal L}}{pv_r}\right)
.
\label{ddSdr}
\end{eqnarray}
The lower boundary condition $\delta v_r(r_{\rm ns})=0$ is written using Eq.~(\ref{dvr}):
\begin{eqnarray}
\left(
 \delta h
- \frac{i\omega' }{ c^2} \frac{r\delta v_\varphi}{ im}
- \frac{\delta k_\theta}{ imc^2}
+ \delta S
 \right)_{\rm ns}
=0.\label{inner_BC}
\end{eqnarray}
Defining $\Delta\zeta$ as the radial displacement of the shock, the boundary conditions at the shock calculated in Appendix~\ref{append_nonax} are expressed as follows:
\begin{eqnarray}
\delta h_{\rm sh}&=&-i\omega'_{\rm sh}\left(1-\frac{v_{\rm sh}}{ v_1}\right)\frac{\Delta \zeta}{ v_{\rm sh}}\label{dhsh0}
,\\
\frac{(r\delta v_\varphi)_{\rm sh}}{ im}&= &
(v_1-v_{\rm sh})\Delta\zeta,
\label{drvphish0}\\
\frac{(\delta k_\theta)_{\rm sh}}{ im}&=&
-\Delta\zeta 
\left\lbrack
\frac{c^2}{\gamma}\nabla S
\right\rbrack_1^2
\label{dkthetash0}
,\\
\frac{\delta S_{\rm sh}}{\gamma}
&=&
(i\omega'_{\rm sh}+\omega_\Phi)
\frac{v_1\Delta \zeta}{ c_{\rm sh}^2} 
\left(1-\frac{v_{\rm sh}}{ v_1}\right)^2
\label{dSsh0},
\end{eqnarray}
where the reference frequency $\omega_\Phi$ defined in Eq.~(26) in Paper~I is generalized to include a centrifugal correction to the gravitational potential:
\begin{eqnarray}
\omega_\Phi\equiv 
-\frac{1}{ v_1}
\frac{
\frac{GM}{ r_{\rm sh}^2}-\frac{L^2}{ r_{\rm sh}^3}
-
2\frac{v_1v_{\rm sh}}{ r_{\rm sh}}
}{
1-\frac{v_{\rm sh}}{ v_1}}
-\frac{c_{\rm sh}^2}{ \gamma v_1}\frac{\left\lbrack \nabla S\right\rbrack^{\rm sh}_1}{ \left(1-\frac{v_{\rm sh}}{ v_1}\right)^2}
.\label{def_om_Phi}
\end{eqnarray}
The derivative of Eq.~(\ref{ddLdr}) is combined with Eq.~(\ref{ddhdr}), using Eqs.~(\ref{ddkdr}-\ref{ddSdr}) and Eq.~(\ref{FSW}), to obtain the following second order equation:
\begin{eqnarray}
\left\lbrack \left(\frac{1-\M^2}{ v_r}\frac{\partial}{\partial r}+\frac{i\omega' }{c^2}\right)^2
+\frac{\omega'^2-\omega_{\rm Lamb}^2}{ v_r^2c^2} 
\right\rbrack
r\delta v_\varphi
=
\nonumber\\
-\frac{1-\M^2}{ v_r}
\left\lbrack
\frac{\partial}{\partial r}\left(\frac{r\delta w_\theta}{v_r}\right)
-im\frac{\gamma-1}{\gamma}\delta\left(\frac{{\cal L}}{pv_r}\right)
\right\rbrack
.
\label{forced_oscillator}
\end{eqnarray}
In the limit of no rotation, this equation is equivalent to Eq.~(27) in Paper~I, where two typos should be corrected. As noted by YF08 in cylindrical geometry, the formal effect of rotation is a transformation of $\omega$ into $\omega'$, and the addition of a centrifugal contribution to the effective gravity in $\omega_\Phi$, and also in the radial profile of $\M,v_r,p,c$ through the stationary flow equation (\ref{eq_euler_stat}).
A more compact formulation of Eq.~(\ref{forced_oscillator}) is obtained using the same radial coordinate $X$ as in Paper~I and defining the perturbative variable $\delta Y$ with the perturbation of specific angular momentum $r\delta v_\varphi/im$ as follows:
\begin{eqnarray}
{\rm d X}&\equiv& \frac{v_r}{ 1-\M^2} {\rm d}r,\label{def_X}\\
 \delta Y& \equiv&
 \frac{r\delta v_\varphi}{im}{\rm e}^{\int_{\rm sh}^r \frac{i\omega'\M^2}{1-\M^2}\frac{{\rm d}r}{v_r}}.\label{def_Y}
\end{eqnarray} 
We note that without rotation the definition (\ref{def_Y}) of $\delta Y$ coincides in the equatorial plane and in the adiabatic approximation with the variable $\delta Y$ defined in Eq.~(59) in Paper~I using $\delta A$ or $\delta f$, according to Eqs.~(\ref{dfk}) and~(\ref{dAvphi}) with $\delta k_\theta=0$.
The compact formulation of Eq.~(\ref{forced_oscillator}) is as follows:
\begin{eqnarray}
\left( \frac{\partial^2}{\partial X^2}
+\frac{\omega'^2-\omega_{\rm Lamb}^2}{ v_r^2c^2} 
\right)
\delta Y
=
\nonumber\\
-{\rm e}^{\int_{\rm sh} \frac{i\omega' }{c^2}{\rm d}X}
\left\lbrack
\frac{\partial}{\partial X}\left(\frac{r\delta w_\theta}{imv_r}\right)
-\frac{1-\M^2}{ v_r}\frac{\gamma-1}{\gamma}\delta\left(\frac{{\cal L}}{pv_r}\right)
\right\rbrack
\label{forced_oscillator_compact}
\end{eqnarray} 
This compact formulation reveals the forced oscillator structure of the perturbative problem, which will be further characterized in the adiabatic approximation in Sect.~\ref{sect_adiabatic}.

\subsection{Numerical and analytical calculation of the eigenfrequencies with rotation and cooling} 

\begin{figure}
\centering
\includegraphics[width=\columnwidth]{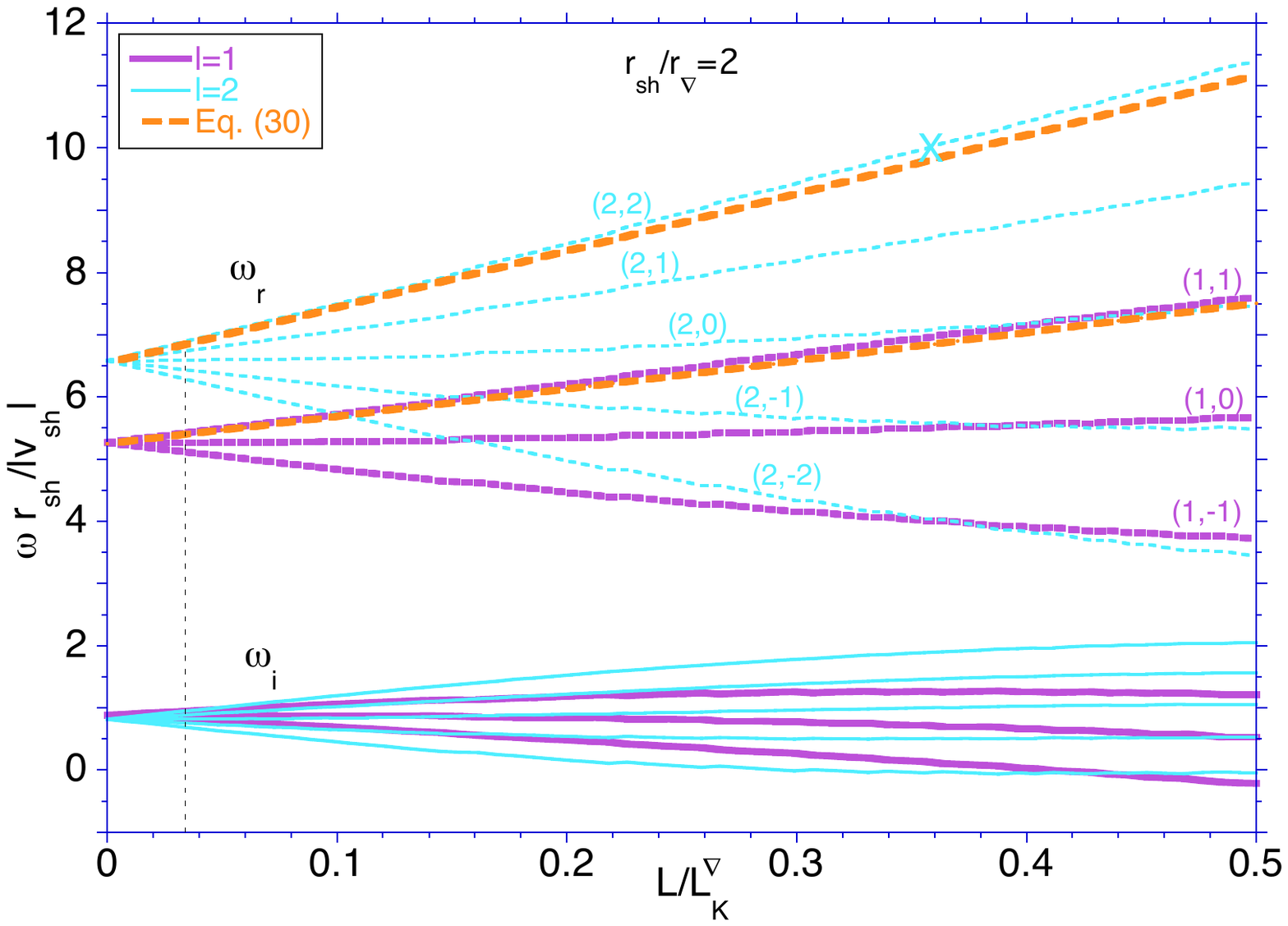}
\caption{Effect of rotation on the eigenfrequencies associated with the spiral SASI modes $\ell=1$ (purple lines) and $\ell=2$ (cyan lines) when the shock radius is $r_{\rm sh}/r_\nabla=2$. The cooling function is defined by $(\alpha,\beta)=(3/2,5/2)$. The Keplerian reference $L_{\rm K}^\nabla$ is defined  at the neutron star surface which coincides with the radius $r_\nabla$ of maximum deceleration. The value of the poloidal and azimuthal wavenumber $(\ell,m)$ is indicated on each frequency branch. The growth rates are shown as solid lines and the frequencies as dashed lines. The analytical fits of $\omega_r$ for the dominant modes $\ell=m=1$ and $\ell=m=2$ (Eq.~\ref{wrfit}) are shown with orange dashed lines. The black vertical dashed line marks the rotation threshold above which the $\ell=m=2$ spiral grows faster than $\ell=m=1$. The cross indicates the frequency threshold above which $r_{\rm co}>r_\nabla$ for the mode $(2,2)$.
}
\label{fig_rot_lm}
\end{figure}

The effect of rotation on the frequency and growth rate of SASI is shown in Fig.~\ref{fig_rot_lm} for each $m$-component of the modes $\ell=1$ and $\ell=2$, for a typical value of the shock radius $r_{\rm sh}/r_\nabla=2$ and a cooling function ${\cal L}$ defined in Eq.~(\ref{def_cool}) with $\alpha=3/2$, $\beta=5/2$. 
The boundary value problem is solved using a shooting method from the shock to the inner boundary, iterating over the value of the complex eigenfrequency $\omega$. 
Fig.~\ref{fig_rot_lm} confirms the results of WFT23 obtained in the hybrid spherical-equatorial approximation: rotation destabilizes the prograde mode of SASI. The main effect of rotation scales approximately linearly with the specific angular momentum $L$ like in cylindrical geometry (YF08). 
This effect is most destabilizing for the prograde spiral $m=+\ell$. 
The fundamental mode $\ell=m=2$ becomes dominant over the mode $\ell=1$ when the rotation rate is increased.\\
\begin{figure}
\centering
\includegraphics[width=\columnwidth]{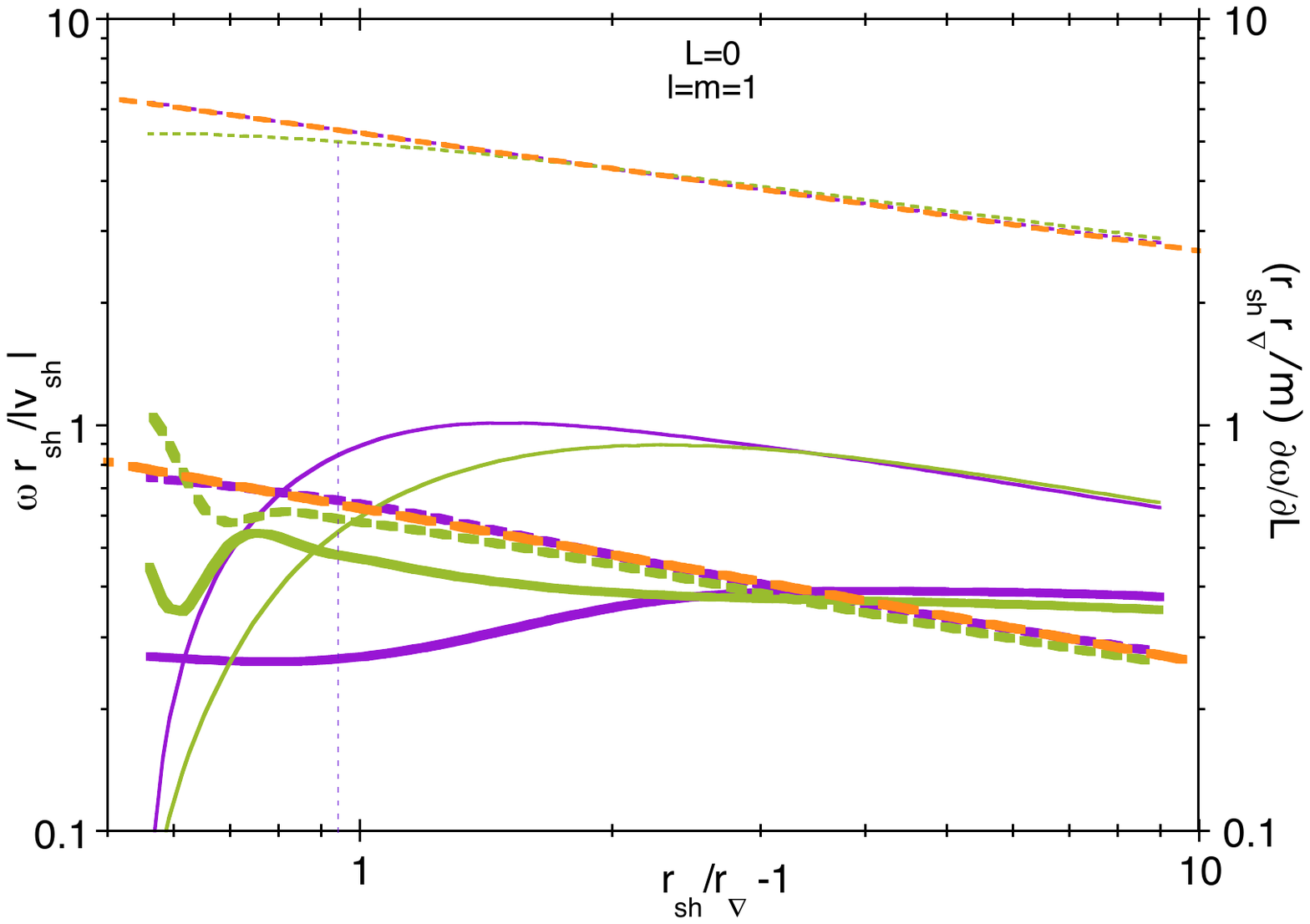}
\includegraphics[width=\columnwidth]{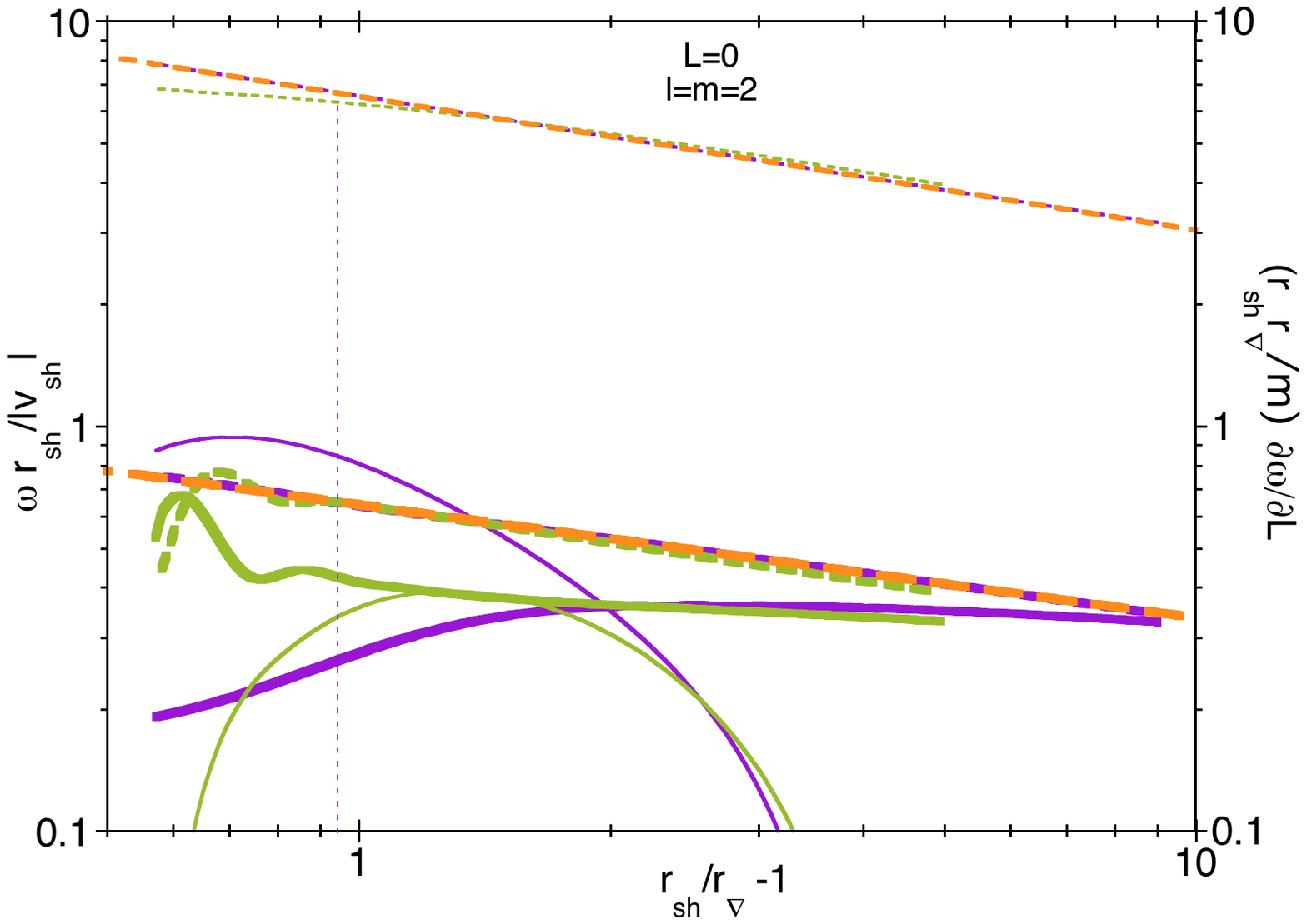}
\caption{Eigenfrequency $\omega$ of the fundamental SASI modes $\ell=m=1$ (upper plot) and $\ell=m=2$ (lower plot) for two different cooling functions with the same shock radius, mass accretion rate without rotation. The growth rate (thin solid lines) and the frequency (thin dashed lines) are measured on the left axis, normalized by $|v_{\rm sh}|/r_{\rm sh}$. The effect of rotation on the growth rate and frequency of the prograde mode $m=\ell$ is measured on the right axis by the derivative $\partial\omega/\partial L$ for $L=0$ (thick solid and dashed lines), normalized by $m/r_{\rm sh}r_\nabla$. The cooling function is either defined by $(\alpha,\beta)=(3/2,5/2)$ (purple lines) or $(\alpha,\beta)=(6,1)$ (green lines). The analytical fits for the frequency of the modes $\ell=m=1$ and $\ell=m=2$ without rotation (Eqs.~\ref{F11},\ref{F22}) and for the derivative $\partial\omega_r/\partial L$ (Eqs.~\ref{G11},\ref{G22}) are shown with orange dashed lines. The purple vertical dashed line marks the threshold below which the $\ell=m=2$ spiral grows faster than $\ell=m=1$ for $(\alpha,\beta)=(3/2,5/2)$.
}
\label{fig_rot_dwdL}
\end{figure}
\begin{figure}
\centering
\includegraphics[width=\columnwidth]{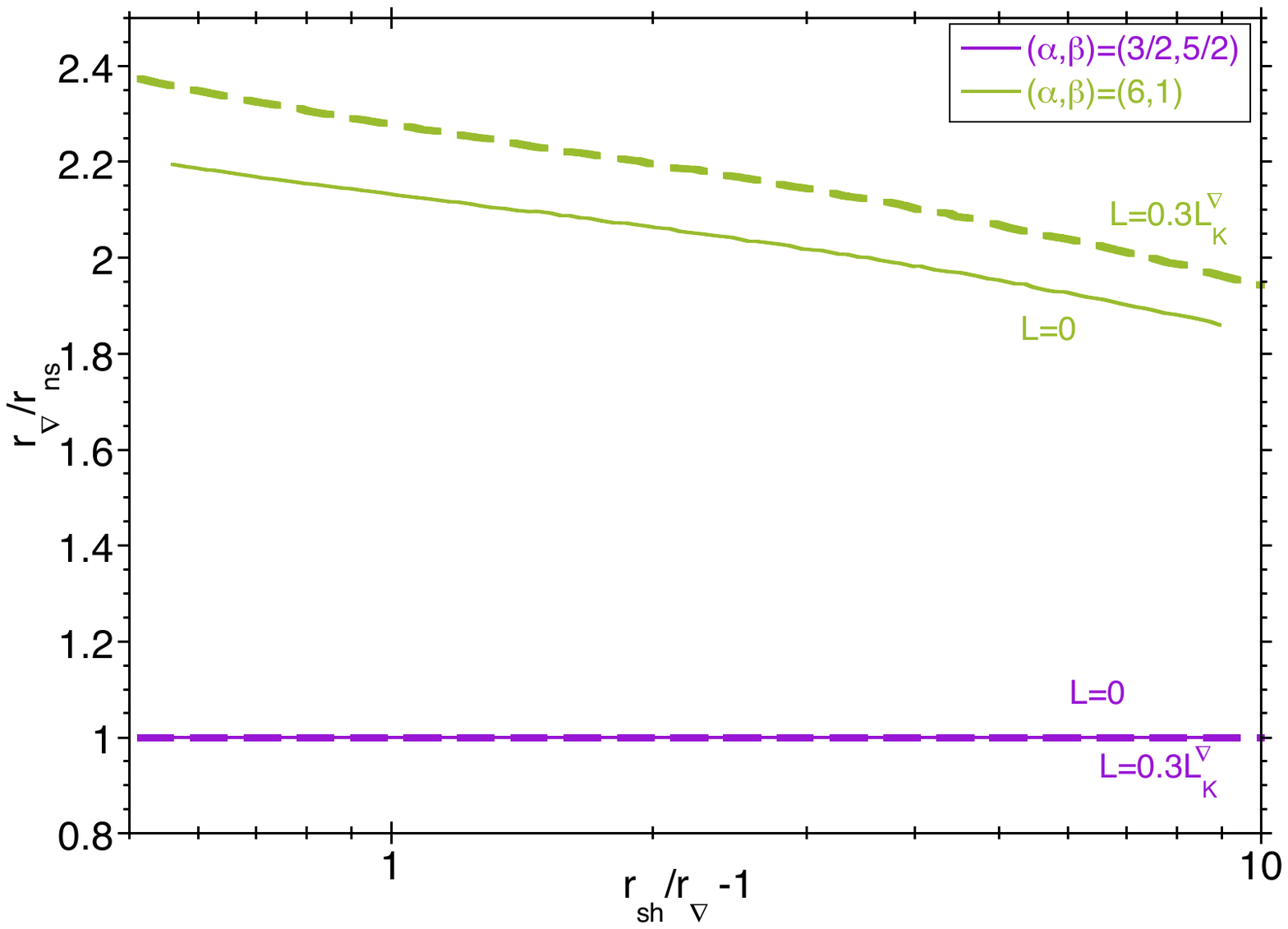}
\caption{Ratio $r_\nabla/r_{\rm ns}$ in the stationary flow depending on the shock radius for $L=0$ (solid lines) and $L=0.3L_{\rm K}^\nabla$ (dashed lines) with the cooling function defined by $(\alpha,\beta)=(3/2,5/2)$ (purple lines) or $(\alpha,\beta)=(6,1)$ (green lines).
}
\label{fig_rshrnab}
\end{figure}
\begin{figure}
\centering
\includegraphics[width=\columnwidth]{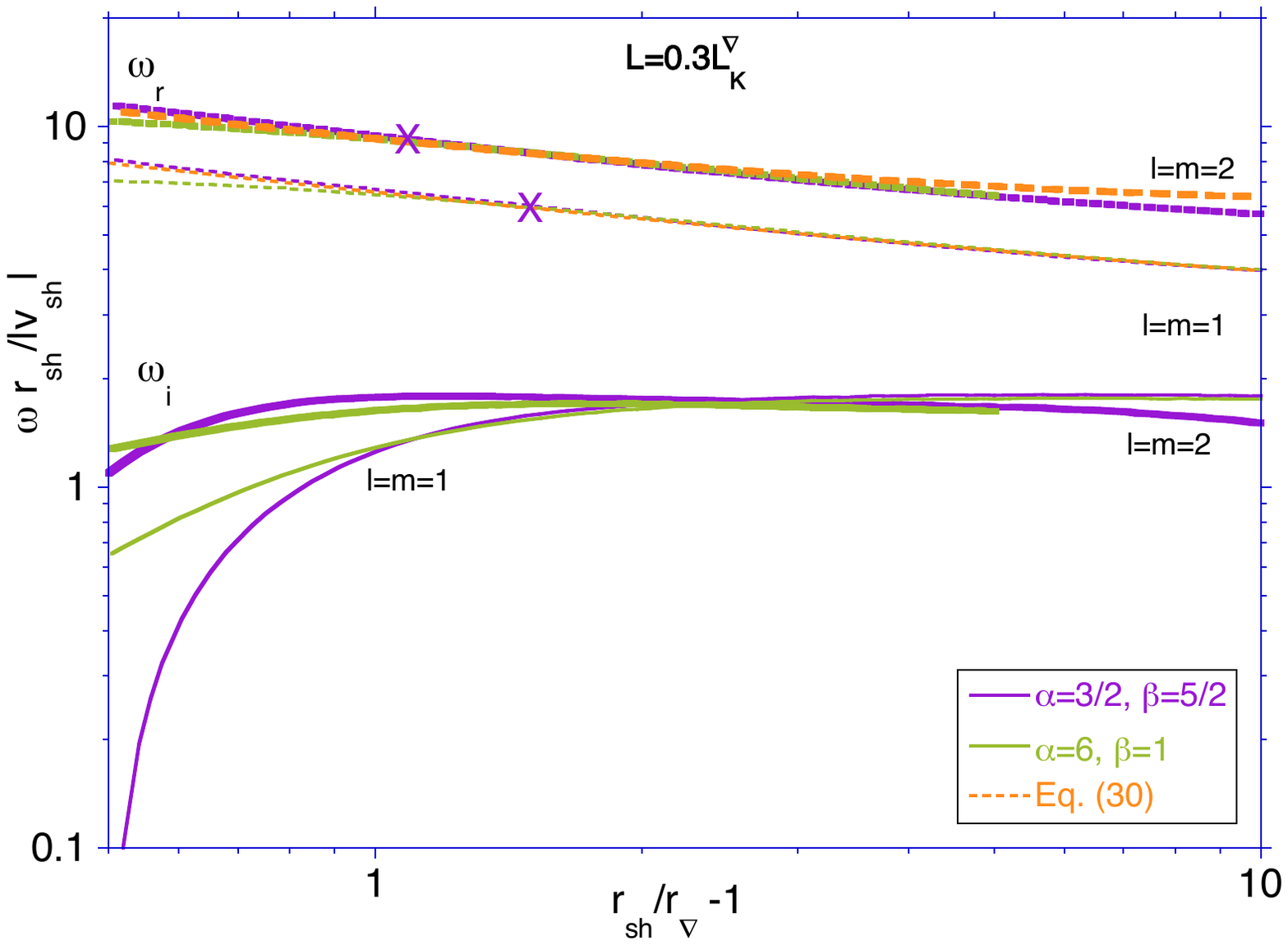}
\caption{Eigenfrequencies of the fundamental SASI modes $\ell=m=1$ (thin lines) and $\ell=m=2$ (thick lines) for two different cooling functions with the same shock radius, mass accretion rate and uniform specific angular momentum in the stationary flow. The magnitude of the specific angular momentum $L$ is chosen as $30\%$ of the Keplerian threshold $L_{\rm K}^\nabla$ defined at $r_\nabla$ for a cooling function defined by $(\alpha,\beta)=(3/2,5/2)$ (magenta lines) or $(6,1)$ (green lines). The growth rates (solid lines) and the frequencies (dashed lines) are normalized by $|v_{\rm sh}|/r_{\rm sh}$. The analytical fits of $\omega_r$ corresponding to Eq.~(\ref{wrfit}) with $m=1$ and $m=2$ are shown with orange dashed lines. The purple cross corresponds to $r_{\rm co}= r_\nabla$. 
}
\label{fig_rot_nabla}
\end{figure}
We note from Fig.~\ref{fig_rot_lm} that the effect of rotation on the eigenfrequency is mainly linear and can be characterized by the slope $\partial\omega/\partial L$ at $L=0$. We calculate in Fig.~\ref{fig_rot_dwdL} the dimensionless derivative $(r_{\rm sh}r_\nabla/m)\partial\omega/\partial L$ for the fundamental mode $\ell=m=2$ and $\ell=m=1$ for $L=0$. We remark that the real part of this quantity depends little on the parameters $(\alpha,\beta)$ of the cooling function for $r_{\rm sh}/r_\nabla>1.5$, despite the significantly smaller radius of the proto-neutron star ($r_{\nabla}/r_{\rm ns}>1.7$) for $\alpha=6$, $\beta=1$ as shown in Fig.~\ref{fig_rshrnab} with a green solid line.
The same is true for the imaginary part of the dimensionless derivative in Fig.~\ref{fig_rot_dwdL} for a more restrictive range $r_{\rm sh}/r_\nabla>3$. 

We can thus define explicit dimensionless functions $F_{\ell,m}(x),G_{\ell,m}(x)$ with $x\equiv r_{\rm sh}/r_\nabla$ in order to approximate the fundamental SASI eigenfrequency $\omega_r$ associated with the spherical harmonic $Y_\ell^m$, in an accretion flow with a specific angular momentum $L$ in the equatorial plane, as follows:
\begin{eqnarray}
\omega_r^{\rm fit}(\ell,m)\sim\frac{|v_{\rm sh}|}{ r_{\rm sh}} F_{\ell,m}\left(\frac{r_{\rm sh}}{ r_\nabla}\right) 
+\frac{mL}{ r_{\rm sh}r_\nabla}G_{\ell,m}\left(\frac{r_{\rm sh}}{ r_\nabla}\right).
\label{wrfit}
\end{eqnarray}
The following approximations of the reference functions $F_{1,1},G_{1,1},F_{2,2},G_{2,2}$ are deduced from Fig.~\ref{fig_rot_dwdL}:
\begin{eqnarray}
F_{1,1}(x)&\equiv&\frac{5.25}{(x-1)^{0.29}}
,\label{F11}\\
G_{1,1}(x)&\equiv&\frac{0.625}{(x-1)^{0.38}},\label{G11}\\
F_{2,2}(x)&\equiv&\frac{6.54}{(x-1)^{0.33}},\label{F22}\\
G_{2,2}(x)&\equiv&\frac{0.642}{(x-1)^{0.28}}.\label{G22}
\end{eqnarray}
The definition (\ref{F11}) of the function $F_{1,1}$ is a simpler form than Eqs.~(29-31) proposed in Paper~I, with comparable accuracy.
The analytical approximation is compared to the exact calculation as orange lines in Figs.~\ref{fig_rot_lm}, \ref{fig_rot_dwdL}, and \ref{fig_rot_nabla}. The scaling factor $v_{\rm sh}/r_{\rm sh}$ can be estimated from $r_{\rm sh}$ and the strength of gravity at the shock assuming free fall, with a quadratic dependence on the rotation rate (Eq.~\ref{eqapeta2}) and a dependence on the compression ratio $v_{\rm sh}/v_1=\rho_1/\rho_{\rm sh}$ in the stationary flow:
\begin{eqnarray}
\frac{|v_{\rm sh}|}{ r_{\rm sh}}&\sim&\left(\frac{2GM}{ r_{\rm sh}^3}\right)^{\frac{1}{2}}
\frac{v_{\rm sh}}{ v_1}
\left\lbrack 1-\left(\frac{L}{ L_{\rm K}^\nabla}\right)^2\frac{r_\nabla}{ 2 r_{\rm sh}}\right\rbrack^{\frac{1}{2}}.
\label{vshrsh}
\end{eqnarray}
The compression ratio depends on photodissociation $\epsilon$ across the shock, which depends on the shock radius $r_{\rm sh}$. A first order estimate of this dependence used in Paper~I (Eqs.~A.5 and A.9) for $\epsilon<0.5$ is 
\begin{eqnarray}
\frac{v_1}{v_{\rm sh}}&\sim& 1+\frac{6}{1-\epsilon},\label{v1vsh}\\
\epsilon&\sim &0.7\left(\frac{r_{\rm sh}}{ 150{\rm km}}\right)\left(\frac{1.3M_\sun}{M_{\rm ns}}\right).\label{epsilon}
\end{eqnarray}

The calculation of SASI eigenfrequencies with $L=0.3L_{\rm K}^\nabla$ in Fig.~\ref{fig_rot_nabla} demonstrates that even with significant rotation the SASI properties are nearly independent of the details of the cooling function when the shock radius is measured in units of $r_{\rm sh}/r_\nabla$ rather than $r_{\rm sh}/r_{\rm ns}$. In a similar way as without rotation (Paper~I), the impact of cooling processes on SASI eigenfrequencies is captured through the parameter $r_\nabla$ defined as the radius of maximum deceleration. The normalized eigenfrequencies are remarkably similar in a flow with $(\alpha,\beta)=(6,1)$ (green lines) with the same angular momentum and the same ratio $r_{\rm sh}/r_\nabla$ (dashed green line in Fig.~\ref{fig_rshrnab}).

\subsection{$\ell=1$ vs $\ell=2$}

\begin{figure}
\centering
\includegraphics[width=\columnwidth]{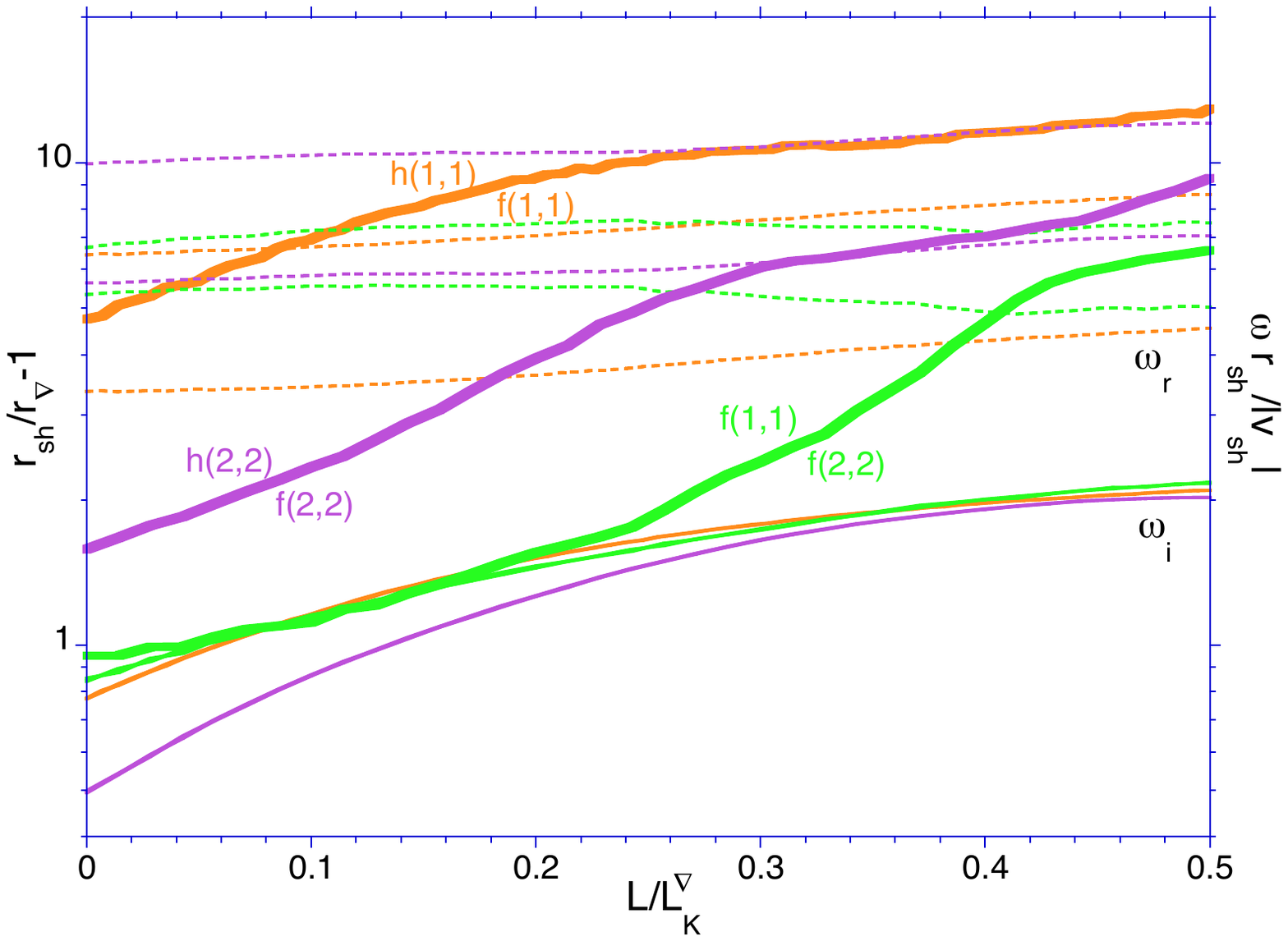}
\caption{Comparing the growth rates of the fundamental spiral SASI modes $\ell=1$ and $\ell=2$ and their first harmonic defines their respective domains of domination which depend on the shock radius $r_{\rm sh}/r_\nabla$ and the rotation rate $L/L_{\rm K}^\nabla$, calculated here with $(\alpha,\beta)=(3/2,5/2)$. The thick solid lines indicate the threshold $r_{\rm sh}/r_\nabla$ (left axis) below which the fundamental mode (noted "f") grows faster than its first harmonic (noted "h"). This transition is shown for the mode $\ell=m=1$ in orange and $\ell=m=2$ in purple. The thick solid green line indicates the threshold $r_{\rm sh}/r_\nabla$ above which the fundamental $\ell=m=1$ spiral grows faster than the fundamental $\ell=m=2$ spiral. For each of these three couples of modes, the growth rates and eigenfrequencies at the transition radius are displayed with thin lines and measured in units of $|v_{\rm sh}|/r_{\rm sh}$ on the right axis. The value of the growth rate is indicated with a thin solid line and the two oscillation frequencies are indicated with thin dashed lines with corresponding colours. }
\label{fig_threshold}
\end{figure}

Distinguishing SASI frequencies associated with $l=1$ and $l=2$ perturbations is important since both can be unstable with comparable growth rates (Figs.~\ref{fig_rot_lm},\ref{fig_rot_nabla}) and may affect the multi-messenger signals. 

Although gravitational waves are mainly produced by $l=2$ perturbations, the non-linear evolution of the mode $\ell=1$ can be a major contributor if the azimuthal density profile steepens. 
The ratio of the $\ell=2$ and $\ell=1$ fundamental frequencies depends on the rotation rate. It is approximated as follows:
\begin{eqnarray}
\frac{\omega_r(2,2)}{\omega_r(1,1)}
=
\frac{F_{2,2}+\frac{2L}{r_\nabla |v_{\rm sh}|}G_{2,2}
}{
F_{1,1}+\frac{L}{ r_\nabla |v_{\rm sh}|}G_{1,1}},\label{ratio12}
\end{eqnarray}
with
\begin{eqnarray}
\frac{L}{ r_\nabla |v_{\rm sh}|}
&= &
\frac{1}{\eta}\frac{L}{ L_{\rm K}^\nabla} \frac{v_1}{v_{\rm sh}}
\left(
\frac{r_{\rm sh}}{ 2r_\nabla}
\right)^{\frac{1}{2}}.
\label{LKnabla}
\end{eqnarray}
We note in Eq.~(\ref{ratio12}) and in Fig.~\ref{fig_rot_nabla} that the frequency of the fundamental mode $\ell=2$ exceeds the frequency of the fundamental mode $\ell=1$ by $20$ to $40\%$ without rotation, and up to a factor $2.3$ with rotation.\\
The measure of both $\omega_r(2,2)$ and $\omega_r(1,1)$ can in principle determine both $r_{\rm sh}/r_\nabla$ and $L/L_{\rm K}^\nabla$.

The thick sold lines in Fig.~\ref{fig_threshold} indicate a transition from a dominant two-arm spiral ($m=2$) to a dominant one-arm spiral ($m=1$) when the ratio $r_{\rm sh}/r_\nabla$ increases. A transition to a higher frequency is also expected for each mode $\ell=1,2$ when the ratio $r_{\rm sh}/r_\nabla$ increases, as the fundamental $m=+\ell$ spiral is dominated by its first overtone. 

One must bear in mind that the growth rate of SASI is sensitive to the specificities of non-adiabatic processes for $r_{\rm sh}<3r_\nabla$ without rotation, and for $r_{\rm sh}<2r_\nabla$ with $L=0.3L_{\rm K}$, thus limiting the predictive power of Fig.~\ref{fig_threshold} which is calculated for the idealized cooling function $\alpha=3/2,\beta=5/2$ and limited to the linear phase of the instability. 

\subsection{Formal comparison with other frequency estimates of SASI-induced oscillations}

The only analytical formulae proposed so far do not include the effect of rotation. 
The formula proposed by \cite{Muller_Janka2014} for the modulation frequency of the neutrino flux by the SASI mode $\ell=1$ is  
\begin{eqnarray}
^1f_{\rm M14}\equiv  \frac{52.6{\rm Hz}}{\log(r_{\rm sh}/r_{\rm ns})}\left(\frac{100{\rm km}}{ r_{\rm sh}}\right)^\frac{3}{2}.\label{MJ14}
\end{eqnarray}
Its comparison to Eq.~(\ref{F11}) is equivalent to Fig.~5 in Paper~I, showing an improved accuracy from $\sim26\%$ to $\sim10\%$.
The value $2\times (^1f_{\rm M14})$ was used by \cite{Powell_Muller2021} to interpret the frequency of gravitational waves as a non linear harmonic of the SASI mode $\ell=1$ in their model ${\rm z85}\_{\rm SFHx}$.
The 'universal' formula proposed by \cite{Torres2021} for the fundamental mode of gravitational waves associated with the post-shock cavity is:
\begin{eqnarray}
^2f_{\rm T21} &\equiv &100{\rm Hz}\left(\frac{M}{ M_{\sun}}\right)^{\frac{1}{2}}\left(\frac{100{\rm km}}{ r_{\rm sh}}\right)^\frac{3}{ 2}\nonumber\\
&&\times\left\lbrace 1.41-0.0423\left(\frac{M}{ M_{\sun}}\right)^{\frac{1}{2}}\left(\frac{100{\rm km}}{ r_{\rm sh}}\right)^\frac{3}{ 2}\right\rbrace.\label{TF21}
\end{eqnarray}
Eq.~(\ref{wrfit}) with Eqs.~(\ref{F11}-\ref{G22}) can be rewritten in a similar format using the normalization (\ref{LKnabla}) of $L$ by $L_{\rm K}^\nabla$ with $\eta\sim1$:
\begin{eqnarray}
^1f&=& 
\frac{7v_{\rm sh}}{v_1}
\left(\frac{M}{ M_{\sun}}\right)^{\frac{1}{2}}
\left(\frac{100{\rm km}}{ r_{\rm sh}}\right)^\frac{3}{ 2}
\times\nonumber\\
&&
\left\lbrack
\frac{61.6{\rm Hz}}{ \left(\frac{r_{\rm sh}}{ r_\nabla}-1\right)^{0.29}}
+
\frac{7.34{\rm Hz}}
{\left(\frac{r_{\rm sh}}{ r_\nabla}-1\right)^{0.38}}
\frac{L}{L_{\rm K}^\nabla}
\left(\frac{r_{\rm sh}}{ 2r_\nabla}\right)^\frac{1}{2}
\right\rbrack
,\label{freqF11}\\
^2f&=&
\frac{7v_{\rm sh}}{ v_1}\left(\frac{M}{ M_{\sun}}\right)^{\frac{1}{2}}
\left(\frac{100{\rm km}}{ r_{\rm sh}}\right)^\frac{3}{ 2}
\times\nonumber\\
&&
\left\lbrack
\frac{76.8{\rm Hz}}{ \left(\frac{r_{\rm sh}}{ r_\nabla}-1\right)^{0.33}}
+
\frac{15.1{\rm Hz}}
{\left(\frac{r_{\rm sh}}{ r_\nabla}-1\right)^{0.28}}
\frac{L}{L_{\rm K}^\nabla}
\left(\frac{r_{\rm sh}}{ 2r_\nabla}\right)^\frac{1}{2}
\right\rbrack
\label{freqF22}
.
\end{eqnarray}
In Eqs.~(\ref{freqF11}) and (\ref{freqF22}), the frequency increase due to rotation is more pronounced for the mode $\ell=2$ than for $\ell=1$ and increases with $r_{\rm sh}$.\\
We note that $^1f_{\rm M14}$, $^2f_{\rm T21}$, $^1f$ and $^2f$ share the main dependence on $r_{\rm sh}^{-3/2}$, but significantly differ in the scaling factor and the deviation from this power law. 
Eq.~(\ref{TF21}) deviates only slightly from a power law, increasing by less than $10\%$ as $r_{\rm sh}$ increases in the range $50$ to $200$km. By contrast, the factor $1/\log(r_{\rm sh}/r_{\rm ns})$ in Eq.~(\ref{MJ14}) is a strongly decreasing functions of $r_{\rm sh}$. The same is true of the factor involving $r_{\rm sh}/r_\nabla-1$ at the denominator in Eqs.~(\ref{freqF11}) and (\ref{freqF22}). The $r_{\rm sh}$-dependent effect of dissociation on compressibility (Eqs.~\ref{v1vsh}-\ref{epsilon}) also contributes to this decrease through the factor $\v2/v_1$. 
Numerical simulations of core-collapse displaying clear SASI signatures in neutrinos and gravitational waves, such as \cite{Powell_Muller2021} without rotation, will be used to assess the domain of validity of these formulae (Moreau et al., in prep).

\section{Adiabatic model with rotation\label{sect_adiabatic}} 

\subsection{Numerical calculation of the eigenfrequencies}

The adiabatic model of Paper~I is extended to the equatorial plane of rotating flows in order to understand the contribution of adiabatic processes to the destabilization of the prograde mode. The adiabatic hypothesis is chosen for the sake of analytical simplicity since it reduces the order of the differential system from 4 to 2. From Eqs.~(\ref{ddkdr}) and (\ref{dkthetash0}) with ${\cal L}=0$ we conclude that $\delta k_\theta$ is uniformly zero in an adiabatic flow,
and the entropy perturbations is simply integrated as follows:
\begin{eqnarray}
\delta S=\delta S_{\rm sh}{\rm e}^{\int_{\rm sh} \frac{i\omega'}{v_r}{\rm d}r}.
\end{eqnarray}
From Eq.~(\ref{FSW}) we deduce the same relation between the baroclinic production of vorticity and the advection of entropy as without rotation (Eq.~43 in Paper~I):
\begin{eqnarray}
\frac{r\delta w_\theta}{v_r}&=&-im\frac{\delta S}{ \gamma\M^2},\\
&=&
\left(\frac{r\delta w}{ v}\right)_{\rm sh}
\frac{\M_{\rm sh}^2}{\M^2}
{\rm e}^{\int_{\rm sh} \frac{i\omega'}{v_r}{\rm d}r}
\label{baroclinic_vort}.
\end{eqnarray}
The boundary condition at the shock for non-axisymmetric equatorial perturbations $m\ne0$ are deduced from Eqs.~(\ref{dhsh0}), (\ref{drvphish0}) and (\ref{dSsh0}) with $\nabla S=0$.
Vorticity perturbations are generated at the shock with an amplitude deduced from Eqs.~(\ref{baroclinic_vort}), (\ref{drvphish0}) and (\ref{dSsh0}):
\begin{eqnarray}
\left(\frac{r\delta w_\theta}{v}\right)_{\rm sh}
=
-\frac{1}{ \v2^2}\left(1-\frac{\v2}{ v_{1}}\right)(i\omega'_{\rm sh}+\omega_\Phi)
(r\delta v_\varphi)_{\rm sh}
\label{Ssh_SW}.
\end{eqnarray} 
We use the same adiabatic boundary condition $\delta v_r=0$ as in Paper~I, deduced from Eq.~(\ref{inner_BC}) with $\delta k_\theta=0$.
\begin{figure}
\centering
\includegraphics[width=\columnwidth]{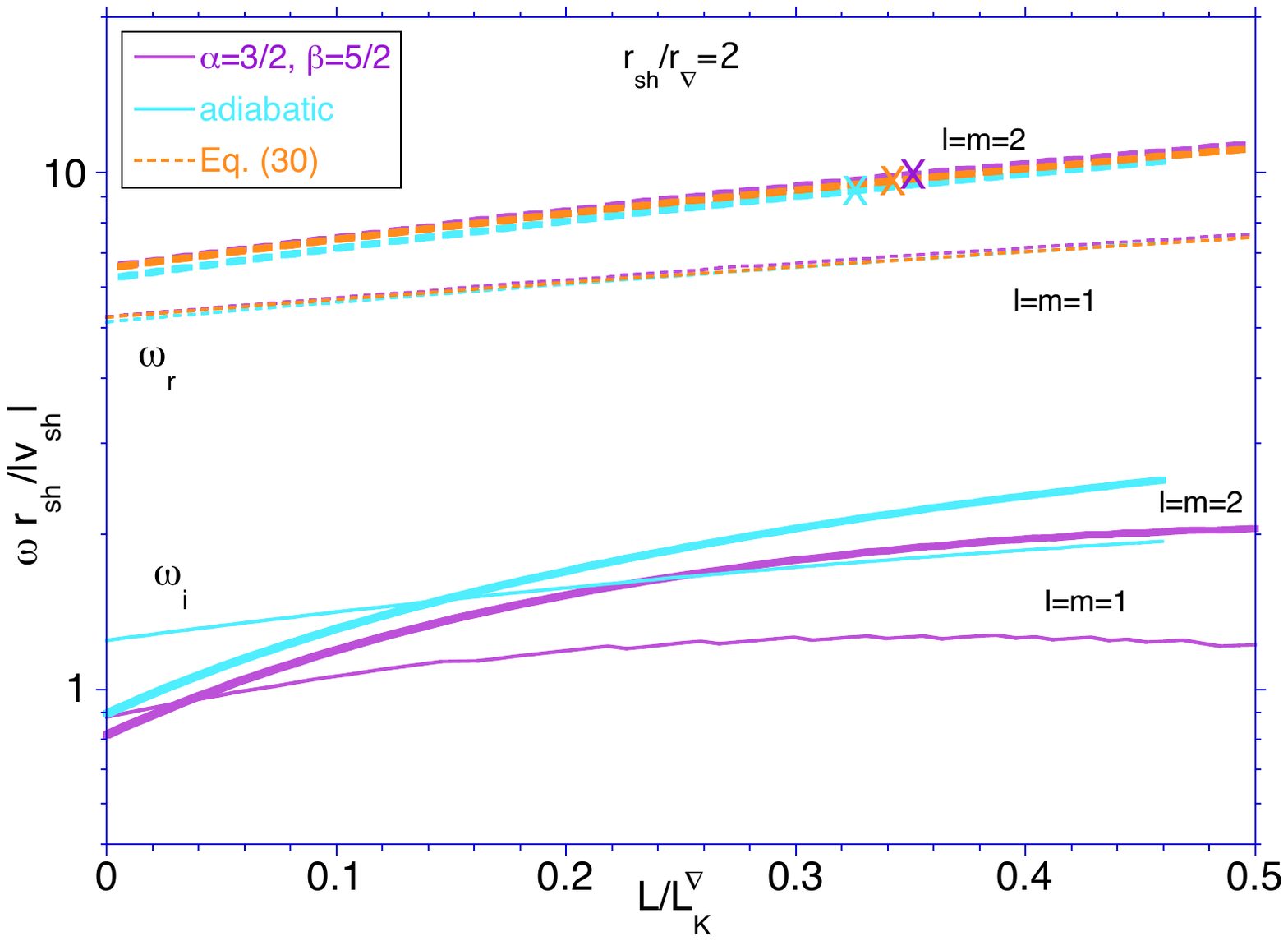}
\includegraphics[width=\columnwidth]{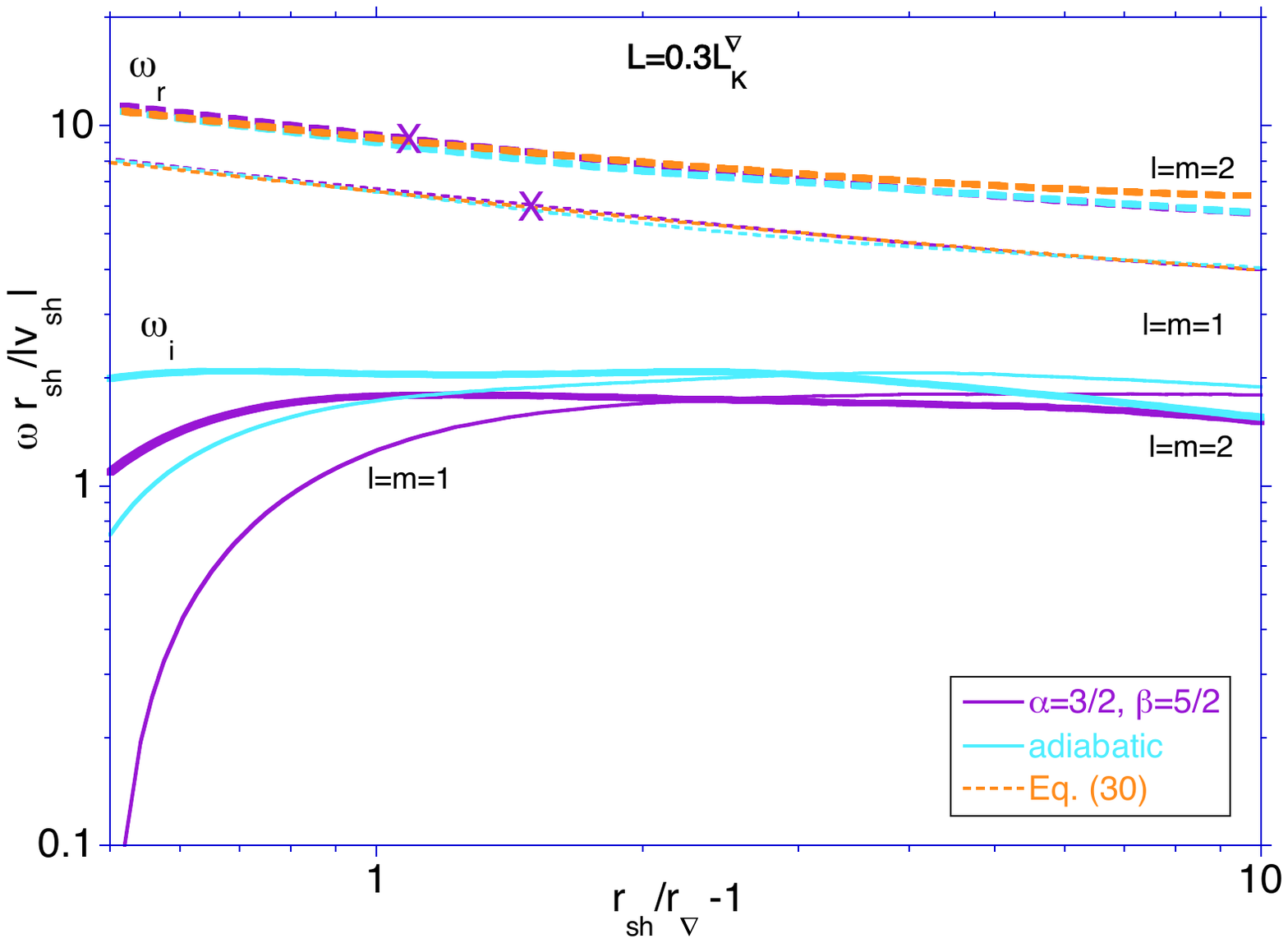}
\caption{The eigenfrequency of the prograde mode $\ell=m=1$ (thin lines) and $\ell=m=2$ (thick lines) is calculated for different rotation rates when the shock distance is $r_{\rm sh}/r_\nabla=2$ (upper plot), and varying $r_{\rm sh}/r_\nabla$ in a collapsing rotating core with 30$\%$ of the Keplerian angular velocity at the inner boundary (bottom plot). The calculation in the adiabatic approximation (cyan lines) is compared to the calculation with $(\alpha,\beta)=(3/2,5/2)$ (purple lines).  The analytical fit of $\omega_r$ (Eq.~\ref{wrfit}) is shown with orange lines.}
\label{fig_cooling}
\end{figure}
\begin{figure}
\centering
\includegraphics[width=\columnwidth]{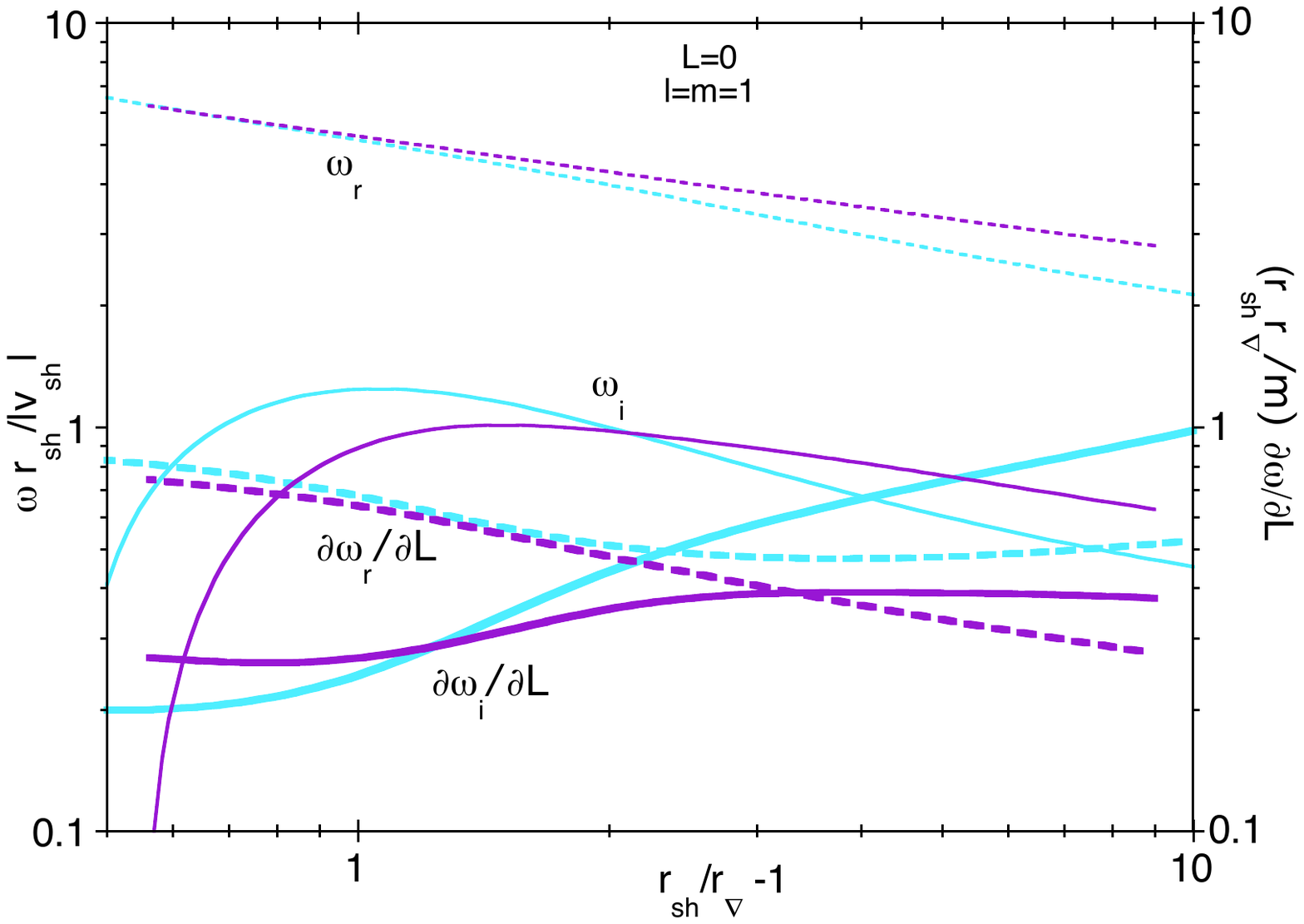}
\includegraphics[width=\columnwidth]{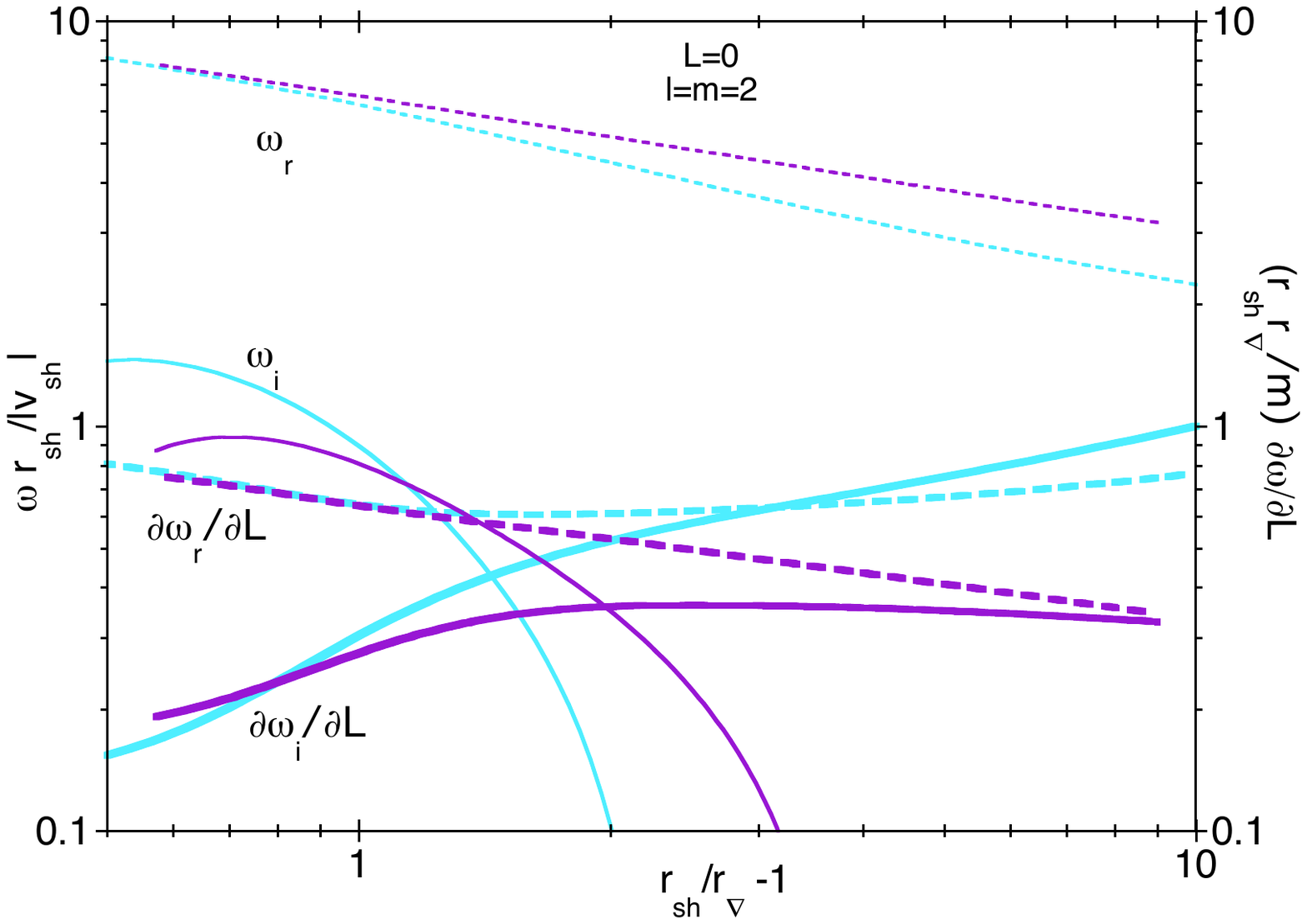}
\caption{The dimensionless measure of the eigenfrequency $\omega r_{\rm sh}/|v_{\rm sh}|$ (left axis) is displayed with thin lines. The dimensionless measure of the derivative $r_{\rm sh}r_\nabla /m \partial \omega/\partial L$ (right axis) is displayed with thick lines. Both are calculated without rotation in the adiabatic approximation (cyan lines) for the modes $\ell=m=1$ (upper plot) and $\ell=m=2$ (lower plot), and compared to the calculation of Fig.~\ref{fig_rot_dwdL} with cooling (purple lines). The real part is shown with dashed lines, and the imaginary part with solid lines. }
\label{fig_dw_r}
\end{figure}
The eigenfrequencies corresponding to the adiabatic system are solved numerically and compared to the eigenfrequencies of the non-adiabatic formulation in Fig.~\ref{fig_cooling} for different rotation rates with $r_{\rm sh}/r_\nabla=2$. 

We note in Fig.~\ref{fig_dw_r} that the oscillation frequencies and their derivative $\partial\omega_r/\partial L$ calculated in the adiabatic approximation are surprisingly close to the values obtained with cooling $(\alpha,\beta)=(3/2,5/2)$ for a small shock radius. The accuracy is better than $10\%$ for $r_{\rm sh}<3r_\nabla$ despite the dominant role of non adiabatic cooling close to the inner boundary.
The adiabatic approximation systematically overestimates $\omega_r$ and $\partial\omega_r/\partial L$ for larger shock radii.

\subsection{Acoustic oscillations forced by the advection of entropy and vorticity}

In the adiabatic approximation the differential equation (\ref{forced_oscillator_compact}) defining the eigenfrequencies is reduced to:
\begin{eqnarray}
\frac{\partial^2 \delta Y}{\partial X^2}+
(\omega'^2-\omega_{\rm Lamb}^2)\frac{\delta Y}{ v_r^2c^2}
=\delta Y_{\rm sh}{\cal F},\label{forced_oscillator_adiab}
\end{eqnarray} 
The description of this system as a "self-forced" oscillation refers to the amplitude of the forcing being proportional to the amplitude of the function $\delta Y$ at the boundary. The radial profile of the coupling term ${\cal F}$ is defined by the radial profile of the vorticity perturbation $\delta w_\theta$, using Eqs.~(\ref{baroclinic_vort}) and (\ref{Ssh_SW}):
\begin{eqnarray}
{\cal F}=
\left(1-\frac{\v2}{ v_{1}}\right)\frac{i\omega'_{\rm sh}+\omega_\Phi}{c_{\rm sh}^2}
{\rm e}^{\int_{\rm sh} \frac{i\omega' }{c^2}{\rm dX}}
\frac{\partial}{\partial X}
\left(\frac{{\rm e}^{\int_{\rm sh} \frac{i\omega' }{v}{\rm dr}}}{ \M^2}\right)
.\label{def_calF}
\end{eqnarray} 
It is thus defined by the radial profile of $1/\M^2$ modulated with the phase oscillations of advected perturbations.\\
As in Paper~I the efficiency of the coupling in a forced oscillator is expected to depend on both the magnitude of the forcing and the phase match between the forcing term and the oscillator.
The quantitative evaluation of this coupling efficiency is analyzed in the next Sect.~\ref{sec_integral_eq} using a classical resolution of the forced oscillator with Green functions. 

\subsection{Integral equation defining the eigenfrequencies\label{sec_integral_eq}}

\begin{figure*}
\centering
\includegraphics[width=\textwidth]{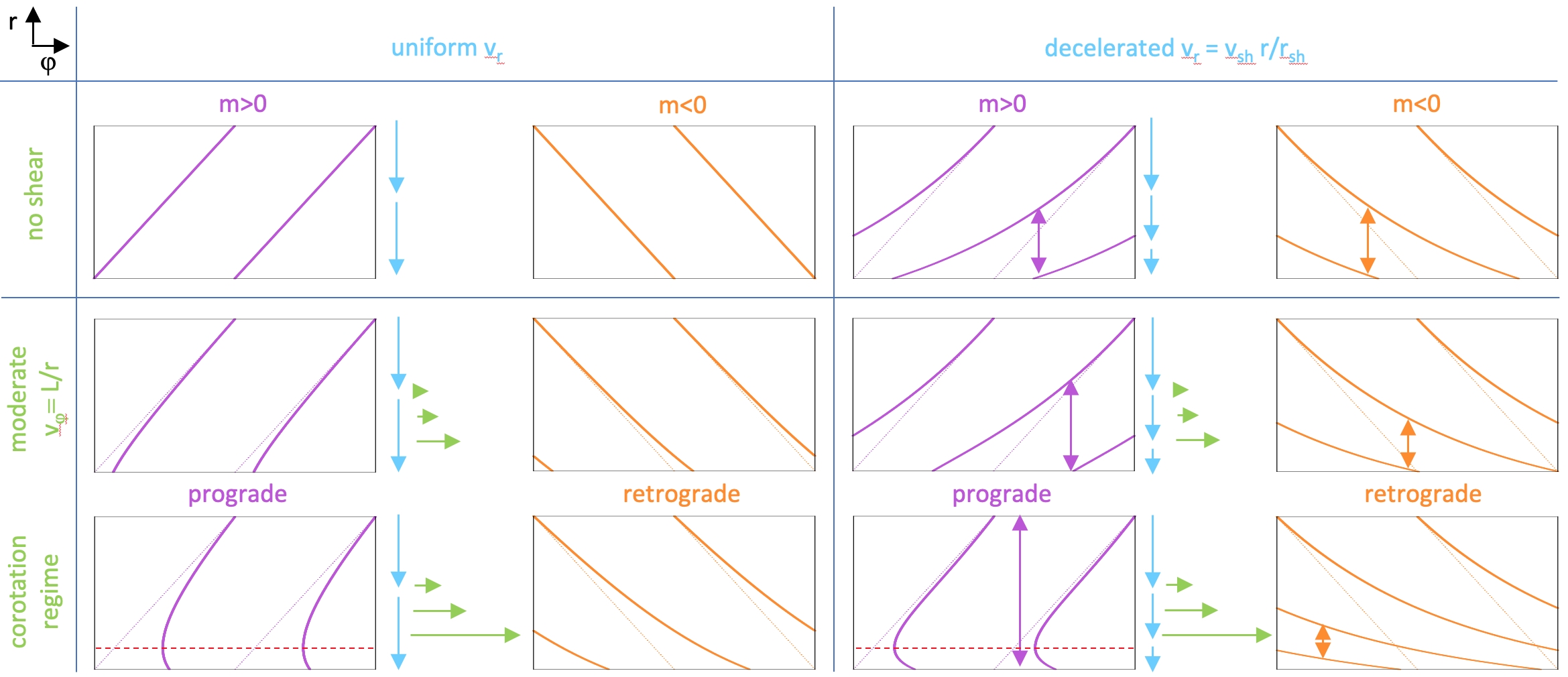}
\caption{The phase $\varphi_{\rm adv}(r)$ of advected perturbations with positive (purple) or negative (orange) azimuthal wavenumber $m$ is affected by both the radial and azimuthal velocity fields $(v_r,v_\varphi)$, illustrated by blue and green arrows. For reference, the first row illustrates the iso-phase lines for radial advection without rotation and the first two columns illustrate the effect of a uniform radial velocity. The second row illustrates the effect of a moderate specific angular momentum, and the third row involves a corotation radius visible as a dashed red line where $\partial \varphi_{\rm adv}/\partial r=0$.
The radial deceleration $v_r\propto r$ illustrated by the right columns decreases the radial wavelength of advected perturbations. The azimuthal shear decreases the radial wavelength of advected perturbation associated with the retrograde mode $m<0$, and increases those associated with the prograde mode $m>0$. The combined effect of deceleration and shear on the phase $\varphi_{\rm adv}(r)$ is approximated by Eq.~(\ref{approx_Psi}).}
\label{fig_decele_shear}
\end{figure*}
We define $Y_0$ as the solution of the homogeneous equation associated with Eq.~(\ref{forced_oscillator_adiab}) with $\delta S=0$ and $\delta K=0$, and satisfying the inner boundary condition (\ref{lowerBCY}):
\begin{eqnarray}
\frac{\partial^2 Y_0}{\partial X^2}+
(\omega'^2-\omega_{\rm Lamb}^2)\frac{Y_0}{ v_r^2c^2}
=0,\label{homogeneous_Y0}\\
\left(\frac{\partial Y_0}{\partial X}\right)_{\rm ns}=\frac{i\omega'_{\rm ns}}{  c_{\rm ns}^2} Y_0^{\rm ns}.\label{homogeneous_LBC}
\end{eqnarray} 
The integral equation defining the eigenfrequencies, derived in Appendix~\ref{integral_dispersion}, is the same as obtained without rotation (Eqs.~63-63 in Paper~I, correcting a sign typo in Eq.~64), except for replacing $\omega$ by $\omega'$ to account for the effect of rotation. This integral equation is further simplified in Appendix~\ref{integral_dispersion} as follows:
\begin{eqnarray}
b_1Y_0^{\rm sh}-
\frac{\v2}{ i\omega'_{\rm sh}}
(1-\M_{\rm sh}^2)b_2  \left(\frac{\partial Y_0}{\partial r}\right)_{\rm sh}
=
\nonumber\\
-\M_{\rm sh}^2\int_{\rm ns}^{\rm sh} 
\frac{1-\M^2}{\M^2}
\frac{\partial Y_0}{\partial r}
{\rm e}^{\int_{\rm sh} \frac{i\omega' }{1-\M^2}\frac{{\rm d}r}{v_r}}
{\rm d}r,\label{single_disp}
\end{eqnarray}
with $b_1,b_2$ defined by:
\begin{eqnarray}
b_1
&\equiv&
\gamma\M_{\rm sh}^2
-\frac{\v2}{ v_1}
b_2
\label{def_b1},\\
b_2&\equiv&
\frac{i\omega'_{\rm sh}}{ i\omega'_{\rm sh}+\omega_\Phi}
\frac{
1
}{
1-\frac{\v2}{ v_{1}}
}
.\label{def_b2}
\end{eqnarray}
We checked numerically that the eigenfrequencies solution of the single equation~(\ref{single_disp}) are strictly the same as obtained in Figs.~\ref{fig_cooling} from the solution of the fourth order adiabatic differential system (\ref{ddhdr}-\ref{ddSdr}) with the boundary conditions defined by Eq.~(\ref{inner_BC}) and Eqs.~(\ref{dhsh0}-\ref{dSsh0}). 

\subsection{Mechanism responsible for the rotational destabilization of SASI}

In the rotation regime $L/L_{\rm K}^{\rm ns}<0.5$ of modest centrifugal effects, rotation does not significantly affect the radial profile of the Mach number. It does affect the radial distribution of the phase and amplitude of the acoustic structure $Y_0$, and the phase distribution of the advected perturbations, through the doppler shifted frequency $\omega'$. The integral formulation (\ref{single_disp}) contains the physical explanation for the destabilizing effect of rotation, both in the regime of moderate rotation and in the corotation regime.
The oscillatory phase $\Psi$ is defined as the real part of the integral term in Eq.~(\ref{single_disp}):
\begin{eqnarray}
\int_{\rm sh}^r \frac{\omega'}{ 1-\M^2}\frac{{\rm d}r}{ v_r}&\equiv&\Psi(r)+i\omega_i \tau_{\rm adv}^{\cal M}(r),\\
\Psi(r)&\equiv&\omega_r \tau_{\rm adv}^{\cal M}(r)-m I_2(r).\label{def_Psi}
\end{eqnarray} 
where $I_2$ is a dimensionless measure of the winding of the flow lines and the timescale $\tau_{\rm adv}^{\cal M}$ is an approximate measure of the advection time:
\begin{eqnarray}
\tau_{\rm adv}^{\cal M}(r)&\equiv& \int_{\rm sh}^r\frac{{\rm d}r}{ (1-\M^2)v_r},\\
I_2(r)&\equiv&\int_{\rm sh}^r  \frac{\Omega{\rm d}r}{ (1-\M^2)v_r}.\label{def_I2}
\end{eqnarray} 
We note that both $I_2$ and $\tau_{\rm adv}^{\cal M}$ are slightly enhanced by a quadratic dependence on the Mach number.\\
In the subsonic limit $\M\ll1$, $\Psi(r)+m\varphi$ coincides with the phase of advected perturbations, illustrated by Fig.~\ref{fig_decele_shear}.
The iso-phase lines $\varphi_{\rm adv}(r)=\varphi_{\rm adv}^{\rm sh}-\Psi(r)/m$ can be calculated explicitly in the approximation where $v_r\sim v_{\rm sh} r/r_{\rm sh}$:
\begin{eqnarray}
\frac{\Psi}{ m}\sim \frac{\omega_r r_{\rm sh}}{ mv_{\rm sh}}\log\frac{r}{ r_{\rm sh}}+\frac{L}{ 2r_{\rm sh}v_{\rm sh}}
\left\lbrack\left(\frac{r_{\rm sh}}{ r}\right)^2-1\right\rbrack.\label{approx_Psi}
\end{eqnarray} 
Figure~\ref{fig_decele_shear} compares iso-phase lines without differential rotation (first row) to their evolution in the regime of moderate rotation (second row) and in the corotation regime (third row): when advected perturbations $\propto \exp\left\lbrack im\varphi_{\rm adv}(r)\right\rbrack$ are sheared by differential rotation their radial wavelength increases or decreases depending on the sign of $m$. The radial wavelength of prograde perturbations ($m>0$) increases, whereas the radial wavelength of retrograde perturbations ($m<0$) decreases. In the regime of moderate rotation the increased radial wavelength of prograde perturbations makes them less prone to phase mixing and favours a stronger advective-acoustic coupling by the flow gradients than without rotation. This coupling is even more favoured in the corotation regime, in the region of stationary phase where $\partial \varphi_{\rm adv}/\partial r=0$.

The Doppler shift of the frequency of the prograde mode in the innermost part of the flow reduces the winding of the perturbations in the region of the temperature peak, as illustrated by the first row of Fig.~10 in \cite{Buellet2023}. The
phase mixing described by the criterion $\omega_r\tau_{\nabla}>1$ in \cite{Foglizzo2009} is thus reduced and the advective-acoustic coupling is enhanced, resulting in a larger growth rate.

With an advective-acoustic coupling taking place closer to the adiabatic temperature peak, the effect of the cooling region ignored in the adiabatic model is likely to affect the result of the calculation with cooling. A tentative explanation for the asymptotically smaller value of $\partial \omega_i/\partial L$ in the solution $\alpha=3/2,\beta=5/2$ in Fig.~\ref{fig_dw_r} could be a smaller intensity of the de-mixing effect due to the flatter temperature peak in the model with cooling compared to the adiabatic model.

\subsection{Stationary phase approximation with a corotation radius}

\begin{figure}
\centering
\includegraphics[width=\columnwidth]{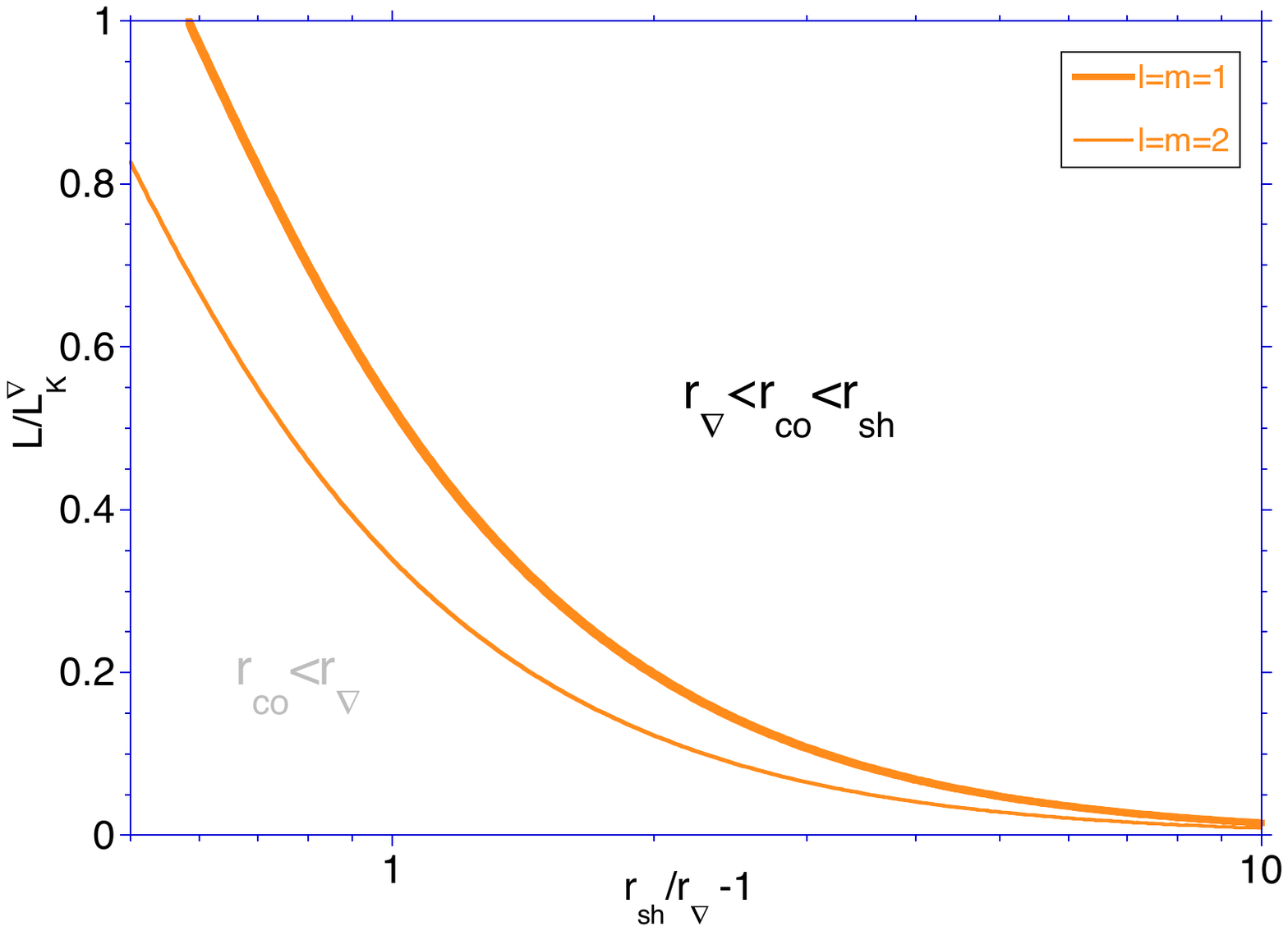}
\caption{Estimated range of shock radius and specific angular momentum allowing for a corotation radius for the fundamental SASI mode $\ell=m=1$ (thick line) and $\ell=m=2$ (thin line), based on the analytical formula (\ref{wrfit}) with Eq.~(\ref{Lrco}).}
\label{fig_rcorot}
\end{figure}

If the specific angular momentum and the shock radius are large enough the phase frequency $\omega/m$ of the eigenmode will coincide with the rotation rate $L/r_{\rm co}^2$ of the fluid at a specific radius defined as the corotation radius $r_{\rm co}$:
\begin{eqnarray}
\omega_r&\equiv&\frac{mL}{ r_{\rm co}^2}\label{defcorot}.
\end{eqnarray} 
The specific angular momentum can be expressed as a fraction of the Keplerian limit $L_{\rm K}$ at the shock:
\begin{eqnarray}
\frac{L}{ r_{\rm sh}|v_{\rm sh}|}
=
\frac{v_1}{ \v2}\frac{1}{2^\frac{1}{ 2}\eta}\frac{L}{ L_{\rm K}^{\rm sh}}.\label{LKepler}
\end{eqnarray} 
Thus
\begin{eqnarray}
\frac{r_{\rm co}}{ r_{\rm sh}}&=&
\left( {\frac{v_1}{\v2}}\frac{M}{2^{\frac{1}{2}}\eta}\frac{L}{ L_{\rm K}^{\rm sh}}
\right)^{\frac{1}{2}}
\left(\frac{|v_{\rm sh}|}{\omega_rr_{\rm sh}}\right)^{\frac{1}{2}}
.
\label{rcosh}
\end{eqnarray} 
The domain of $r_{\rm sh}/r_\nabla$ and $L/L_{\rm K}^\nabla$ allowing for a corotation radius $r_\nabla<r_{\rm co}<r_{\rm sh}$ is estimated using the approximation of $\omega_r$ in Eq.~(\ref{wrfit}) and shown in Fig.~\ref{fig_rcorot}:
\begin{eqnarray}
\left(\frac{L}{  L_{\rm K}^\nabla}\right)^2
=
\frac{
\frac{2r_{\rm sh}}{ r_\nabla}
}{
1
+
\left(
m
\frac{r_{\rm sh}}{ r_\nabla}
{\frac{v_1}{\v2}}
\frac{
\frac{r_{\rm sh}r_\nabla}{ r_{\rm co}^2}
-
G_{\ell,m}
}{
F_{\ell,m}
}
\right)^2
}
.\label{Lrco}
\end{eqnarray} 
The phase $\Psi(r)$ defined by Eq.~(\ref{def_Psi}) is stationary around the corotation radius.
The radial extent $\Delta r$ of the region of stationary phase is characterized using a Taylor expansion. We define the variable $\tau_{\rm ph}(r)$ as follows:
\begin{eqnarray}
\tau_{\rm ph}(r)&\equiv &
\tau_{\rm adv}^{\cal M}(r)
-r^2\int_{\rm sh}^{r} \frac{{\rm dr}}{ (1-\M^2)r^2v_r}.\label{deftau3}
\end{eqnarray} 
We note $\Psi_{\rm co}\equiv\Psi(r_{\rm co})$ and deduce from Eqs.~(\ref{defcorot}) and (\ref{deftau3}):
\begin{eqnarray}
\Psi_{\rm co}=\omega_r\tau_{\rm ph}^{\rm co}.
\end{eqnarray} 
A Taylor expansion of the phase $\Psi$ is performed near the corotation radius:
\begin{eqnarray}
\frac{\partial\Psi}{\partial r}&\equiv&\frac{\omega_r-m\Omega}{(1-\M^2)v_r},\\
\left(\frac{\partial^2\Psi}{\partial r^2}\right)_{\rm co}
&=&
\frac{2mL}{ (1-\M_{\rm co}^2)r_{\rm co}^3 v_{\rm co}}<0,\\
\Psi
&=& \Psi_{\rm co}-\left(\frac{r-r_{\rm co}}{\Delta r}\right)^2
+{\cal O}\left((r-r_{\rm co})^3\right)
,\\
\frac{\Delta r}{ r_{\rm co}} &\equiv& \left\lbrack\frac{(1-\M_{\rm co}^2)|v_{\rm co}|r_{\rm co}}{ mL}\right\rbrack^{\frac{1}{2}},\label{def_deltar}\\
\frac{\Delta r}{ r_{\rm sh}}
&=&(1-\M_{\rm co}^2)^{\frac{1}{2}}
\left({\frac{v_1}{\v2}}\frac{M}{2^{\frac{1}{2}}\eta}\frac{L}{ L_{\rm K}^{\rm sh}}
\right)^\frac{3}{4}\left(\frac{|v_{\rm sh}|}{\omega_rr_{\rm sh}}\right)^\frac{5}{4}.\label{asymptdr}
\end{eqnarray} 
\begin{figure}
\centering
\includegraphics[width=\columnwidth]{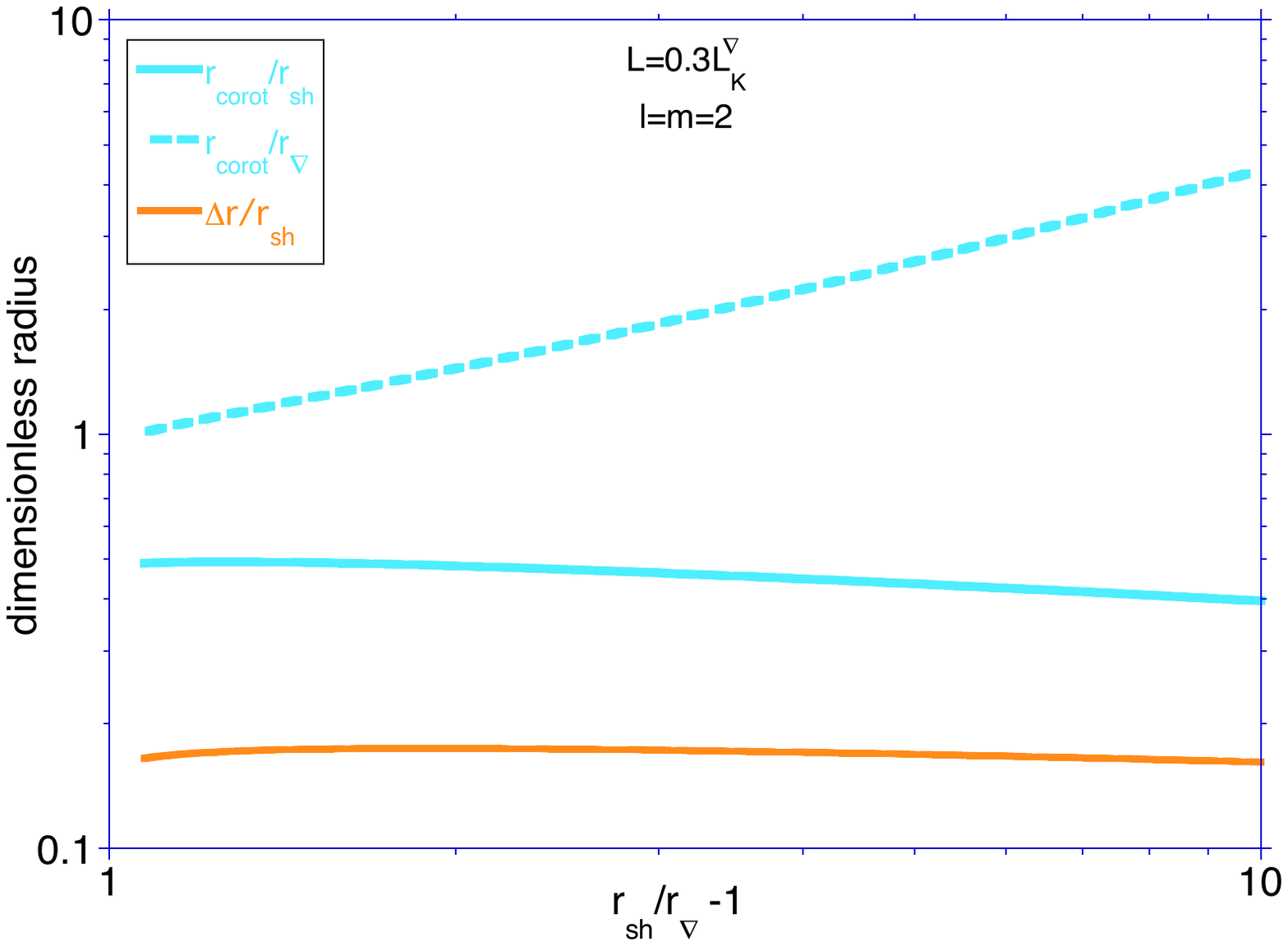}
\includegraphics[width=\columnwidth]{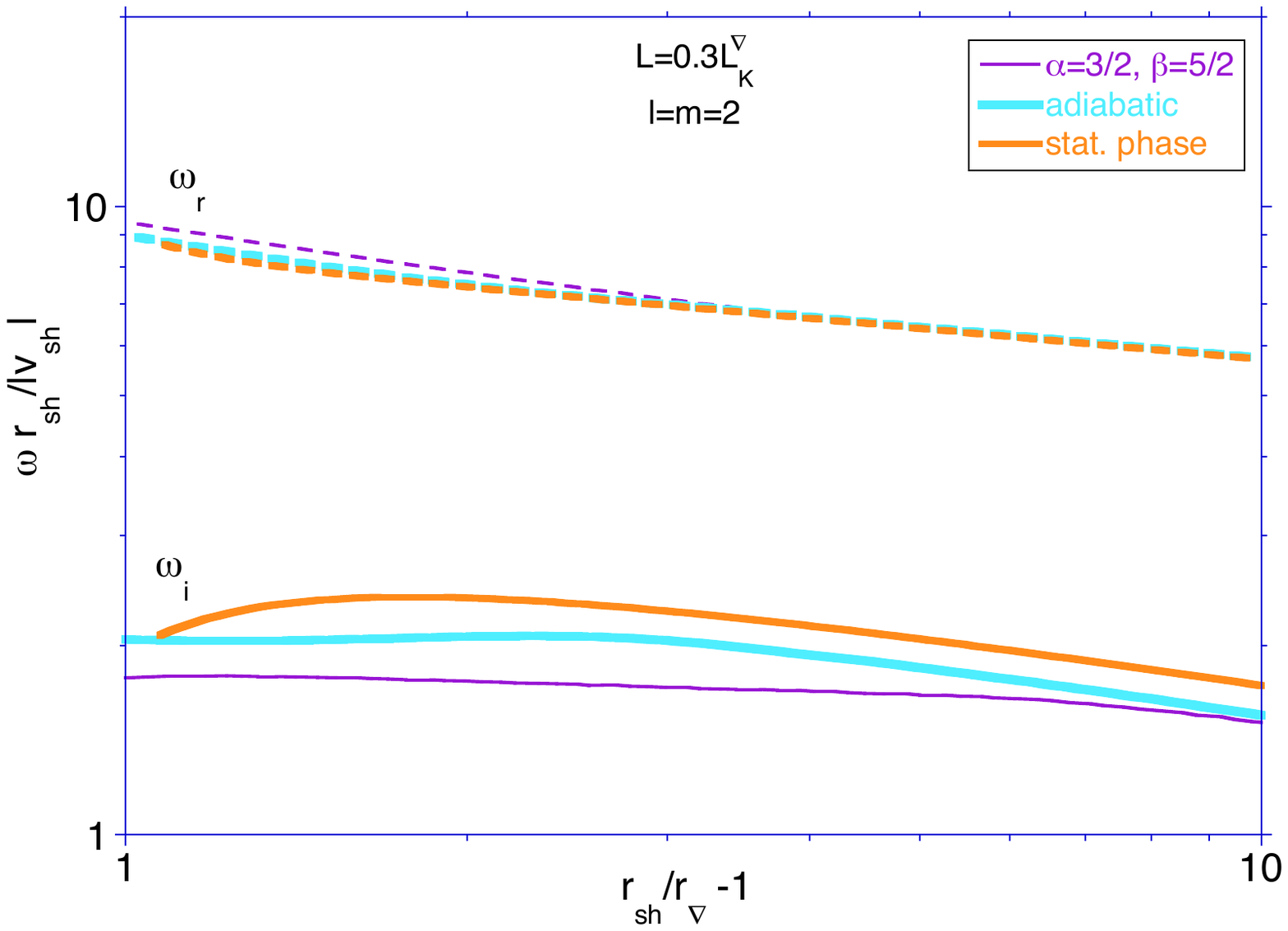}
\caption{In a rotating core with $30\%$ of the Keplerian velocity at the inner boundary. In the upper plot, the corotation radius associated with the mode $l=m=2$ is compared to $r_{\rm sh}$ and $r_{\nabla}$ with cyan lines in the upper plot. The radial extension of the region of stationary phase defined by Eq.~(\ref{def_deltar}) is compared to $r_{\rm sh}$ (orange solid line). In the bottom plot, the frequency (orange dashed line) and growth rate (orange solid line) associated with the mode $l=m=2$ is estimated using the stationary phase approximation (Eq.~\ref{approx_dispers}). The adiabatic calculation (Eq.~\ref{single_disp}) and the calculation with cooling ($\alpha=3/2,\beta=5/2$) shown in Fig.~\ref{fig_cooling} are repeated as blue and purple lines for reference.}
\label{fig_stat}
\end{figure}
We note in Fig.~\ref{fig_stat} that $r_{\rm co}/r_{\rm sh}\sim 0.4-0.5$ and $\Delta r/r_{\rm sh}\sim 0.16-0.18$ weakly depend on the ratio $r_{\rm sh}/r_\nabla$, whereas $r_{\rm co}/r_{\nabla}$ increases nearly linearly with $r_{\rm sh}/r_\nabla$. The oscillation of the phase $\Psi$ between $r_{\rm ns}$ and $r_{\rm co}-\Delta r$ does produce some phase mixing ($(\Psi_{\rm co}-\Psi_{\rm ns})/2\pi\sim 6$), resulting in a negligible contribution of this region to the integral despite the diverging amplitude $\propto 1/\M^2$. The stationary phase approximation is adapted to estimate the integral in this lower region. The modest variation of the phase $\Psi$ between the corotation radius and the shock ($\Psi_{\rm co}/2\pi <1 $) however precludes the stationary phase approximation above the corotation radius, reflecting the fact the the advective-acoustic coupling is radially extended from the corotation radius to the shock. 
Using the stationary phase approximation only below the corotation radius, the integral in Eq.~(\ref{single_disp}) is thus replaced by an integral from $r_{\rm co}$ to $r_{\rm sh}$ as follows:
\begin{eqnarray}
\int_{\rm ns}^{\rm sh} 
\frac{1-\M^2}{\M^2}\frac{\partial Y_0}{\partial r} {\rm e}^{\int_{\rm sh} \frac{i\omega' }{1-\M^2}\frac{{\rm d}r}{v_r}}
{\rm d}r
\nonumber\\
\sim
{\rm e}^{i\Psi_{\rm co}}
\int_{\rm ns}^{\rm sh} 
(1-\M^2)
{{\rm e}^{-\omega_i\tau_{\rm adv}^\frac{\M}(r)}}{\M^2}
\frac{\partial Y_0}{\partial r}
{\rm e}^{-i\left(\frac{r-r_{\rm co}}{\Delta r}\right)^2}
{\rm d}r
,\\
\sim
\frac{\pi^{\frac{1}{2}}}{ 2}
{\rm e}^{i\left(\Psi_{\rm co}-\frac{\pi}{ 4}\right)}
{\rm e}^{-\omega_i\tau_{\rm adv}^{\M}(r_{\rm co})}
\left(
\frac{1-\M^2}{\M^2}
\frac{\partial Y_0}{\partial r}
 \right)_{\rm co}
\Delta r
\nonumber\\
+
\int_{\rm co}^{\rm sh} 
\frac{1-\M^2}{\M^2}\frac{\partial Y_0}{\partial r} {\rm e}^{\int_{\rm sh} \frac{i\omega' }{1-\M^2}\frac{{\rm d}r}{v_r}}
{\rm d}r.
\label{stat_phase}
\end{eqnarray} 
This approximation assumes that the dominant phase oscillation comes from the phase of advected perturbations, modified by differential rotation. It neglects the phase variations associated with the radial structure of the homogeneous solution $Y_0$, which is justified at the low frequency of the fundamental SASI mode. The relation (\ref{single_disp}) defining the eigenfrequencies is thus approximated by:
\begin{eqnarray}
b_1Y_0^{\rm sh}-
\frac{\v2}{ i\omega'_{\rm sh}}
(1-\M_{\rm sh}^2)b_2  \left(\frac{\partial Y_0}{\partial r}\right)_{\rm sh}
=\nonumber\\
\frac{\pi^{\frac{1}{2}}}{ 2}{\rm e}^{i\left(\Psi_{\rm co}-\frac{5\pi}{4}\right)}
{\rm e}^{-\omega_i\tau_{\rm adv}^{\cal M}(r_{\rm co})}
\M_{\rm sh}^2
\left(
\frac{1-\M^2}{\M^2}
\frac{\partial Y_0}{\partial \log r}
 \right)_{\rm co}
\frac{\Delta r}{ r_{\rm co}}
\nonumber\\
-\M_{\rm sh}^2
\int_{\rm co}^{\rm sh} 
\frac{1-\M^2}{\M^2}\frac{\partial Y_0}{\partial r} {\rm e}^{\int_{\rm sh} \frac{i\omega' }{1-\M^2}\frac{{\rm d}r}{v_r}}
{\rm d}r
,\label{approx_dispers}
\end{eqnarray}
where $\tau_{\rm adv}^{\cal M}(r_{\rm co})$ is slightly longer than the advection time from the shock to the corotation point $r_{\rm co}$, and where $r_{\rm co}$ is a function of $\omega_r$ according to Eq.~(\ref{defcorot}) or Eq.~(\ref{rcosh}). 
We note that this formula depends on the flow properties between $r_{\rm co}$ and $r_{\rm sh}$, and is independent of the vicinity of the inner boundary $r_{\rm ns}$ where non adiabatic cooling effects can be important without rotation. This suggests that an analytic formula extending the validity of Eq.~(\ref{wrfit}) to the corotation regime is expected to be formally independent of $r_\nabla$ when $r_{\rm sh}\gg r_{\rm ns}$.

The solution of Eq.~(\ref{approx_dispers}), shown in the bottom plot in Fig.~\ref{fig_stat}, is able to predict the oscillation frequency with a remarkable $<3\%$ accuracy. It overestimates the adiabatic growth rate by $\sim15\%$ or $\sim 0.2|v_{\rm sh}|/r_{\rm sh}$. 

The phase mixing in the region between the neutron star and the corotation radius, demonstrated by the stationary phase approximation, explains the puzzling property that the frequency expressed in units of $|v_{\rm sh}|/r_{\rm sh}$ is little sensitive to the cylindrical or spherical geometry \citep{Walk2023} in the asymptotic limit of a large shock radius: with $r_{\rm co}/r_{\rm sh}\sim 0.4-0.5$ according to the upper plot in Fig.~\ref{fig_stat}, 
the advection and acoustic times in the region $r_{\rm co}<r<r_{\rm sh}$ are not expected to be very sensitive to the geometry. 

For large enough rotation rates, a significant contribution from azimuthal acoustic waves destabilized by a corotation radius is not excluded. Our calculation and physical interpretation open the possibility of a continuous transition between these mechanisms when the specific angular momentum $L$ is increased.

With eigenfrequencies defined in Eq.~(\ref{approx_dispers}) by the region localized near and above the corotation radius, the instability in this regime is less sensitive to non-adiabatic processes of neutrino emission taking place near the neutrinosphere. In the present model neglecting neutrino absorption, the adiabatic approximation is asymptotically accurate for a large shock radius as seen in the bottom plot of Fig.~\ref{fig_stat}.

\section{Conclusions and perspectives}
\label{Sect_discussion}

\subsection{Conclusions}

By taking into account stellar rotation, the present perturbative study incorporates major improvements compared to Paper~I: this is the first perturbative study which takes into account both wavenumbers $\ell,m$, without assuming a cylindrical structure \citep{Yamasaki2008} or neglecting the poloidal derivatives \citep{Walk2023, Buellet2023}.

This study proposes analytical fits of the SASI frequency to interpret its multimessenger signatures with a potentialy larger applicability than existing formulae \citep{Muller_Janka2014,Torres2021}. For each set of wavenumbers $\ell,m$, the physical arameters involved are the shock radius $r_{\rm sh}$, the radius of maximum deceleration $r_\nabla$, the central mass $M_{\rm ns}$, the compression factor $v_1/v_2$ across the shock, the specific angular momentum $L$.

This study also sheds light on the effect of rotation on SASI by focusing on the adiabatic approximation and demonstrating that even with rotation the SASI mechanism can be described as a self-forced oscillator (Eq.~\ref{forced_oscillator}): 
\par(i) a moderate rotation is very efficient at further destabilising the prograde mode by increasing the wavelength of advected perturbations in the deepest regions of the flow and thus decreasing the degree of phase mixing,
\par(ii) if the rotation is fast enough to involve a corotation radius, the dominant region of coupling occurs in the region of corotation and above it, as demonstrated by the stationary phase approximation (Eq.~\ref{approx_dispers} and Fig.~\ref{fig_stat}).

This physical interpretation of the numerical results of YF08, both in the regime of moderate rotation and in the corotation regime, also clarifies the challenging results of WFT23 questioning the continuity of the mechanism between SASI and spiral SASI.

\subsection{Perspectives}

Analytical formulae can improve our ability to identify the physical parameters of the collapsing stellar core in future multimessenger observations of supernovae \citep{Andresen2019,Torres2021}.

The analytical formula describing the eigenfrequency can be tested on non-axisymmetric simulations including rotation in the phase of their dynamical evolution dominated by SASI such as \cite{Iwakami2009a, Iwakami2009b, Iwakami2014, Kazeroni2017, Blondin2017, Summa2018, Andresen2019,Takiwaki2021}.

As already stressed in Paper~I without rotation, the accuracy of the estimation of the SASI oscillation period is limited by the simplicity of the prescription describing partial dissociation at the shock which should also be taken into account within the postshock flow. The relative effect of partial dissociation is expected to be smaller for a small shock radius.   

Another limiting factor is the impact of neutrino absorption in the gain region, neglected in this study, which affects the advection time and the interplay of SASI with the convective instability (\eg \cite{Foglizzo2006,Buellet2023}). Absorption of neutrinos in the gain region is particularly not negligible close to the explosion threshold.

Ideally the interpretation of gravitational waves and neutrinos would require a more global model incorporating SASI, convection in the gain region and in the proto-neutron star, and the low T/W instability \citep{Takiwaki2021,Bugli2023}. This will be the focus of a future study.

\begin{acknowledgements}
It is a pleasure to acknowledge stimulating discussions with J. Guilet, M. Bugli, A. Moreau, H.-T. Janka, B. M\"uller, J. Powell and the LEAK collaboration. This work has received support from the LabEx UnivEarthS, the PNHE/ATPEM and the investment programme “France 2030” as part of the IdEx programme (ANR-18-IDEX-0001) implemented by Université Paris Cité, under which the inIdEx project HERMES is conducted.

\end{acknowledgements}

\bibliographystyle{aa}
\bibliography{references.bib}

\begin{appendix}

\section{Differential system and boundary conditions of perturbed equatorial accretion with rotation}
\label{append_nonax}

The vorticity equation is linearized as follows:
\begin{eqnarray}
-i\omega\frac{\delta w}{\rho}+(v\cdot\nabla)\frac{\delta w}{\rho} = \left(\frac{\delta w}{\rho}\cdot\nabla\right)v\nonumber\\
+\frac{1}{\gamma\rho}\nabla c^2\times\nabla \delta S+\frac{1}{\gamma\rho}\nabla \delta c^2\times\nabla S
.\label{specvortiperturb}
\end{eqnarray}
The perturbed baroclinic terms in Eq.~(\ref{specvortiperturb}) are orthogonal to the radial direction because $\nabla c^2$ and $\nabla S$ in the stationary solution are both radial in the equatorial plane. Using spherical coordinates, the radial vorticity equation is thus simplified into:
\begin{eqnarray}
\frac{\partial}{\partial r}\frac{\delta w_r}{\rho}
=
\frac{i\omega'}{v_r}\frac{\delta w_r}{\rho}
+\frac{\delta w_r}{\rho v_r}\frac{\partial v_r}{\partial r}
\label{dwrdr}.
\end{eqnarray}
The conservation of the tangential component of the velocity across the shock leads to $(\delta w_r)_{\rm sh}=0$ as in Eq.~(B18) in Paper~I, and Eq.~(\ref{dwrdr}) implies $\delta w_r=0$.
Focusing on the mirror symmetric solution such that $\partial\delta w_\theta/\partial\theta=0$ and $\partial\delta w_\varphi/\partial\theta=0$, the transverse vorticity equations in spherical coordinates are multiplied by $\rho r^2 v_r$ and become, in the equatorial plane:
\begin{eqnarray}
\left(\frac{\partial}{\partial r}-\frac{i\omega'}{v_r}\right){rv_r\delta w_\theta}
=
-\frac{im}{\gamma}
\left(
\frac{\partial c^2}{\partial r} \delta S 
-\frac{\partial S}{\partial r}\delta c^2
\right)
,\label{dwthetadr}
\\
\left(\frac{\partial}{\partial r}-\frac{i\omega'}{v_r}\right){rv_r\delta w_\varphi}
=
r^2\delta w_r\frac{\partial\Omega}{\partial r}
+
\frac{1}{\gamma}\frac{\partial}{\partial \theta} 
\left(\frac{\partial c^2}{\partial r}\delta S
-\frac{\partial S}{\partial r}\delta c^2
\right)
.\label{dwphidr}
\end{eqnarray}
The entropy equation is simply
\begin{eqnarray}
\left(\frac{\partial}{\partial r}-\frac{i\omega'}{v_r}\right)\delta S=\delta\left(\frac{{\cal L}}{pv_r}\right).\label{dSdr}
\end{eqnarray}
We define the vector $\delta {\bf k}$ as follows:
\begin{eqnarray}
\delta {\bf k}&\equiv&rv_r\delta {\bf w} -\frac{c^2}{\gamma}{\bf e_r}\times\nabla\delta S,\\
\delta k_r& = &0,\label{defkr}\\
\delta k_\theta& = &rv_r\delta w_\theta+\frac{c^2}{\gamma}\frac{im}{\sin\theta}\delta S,\label{defktheta}\\
\delta k_\varphi&=& rv_r\delta w_\varphi-\frac{c^2}{\gamma}\frac{\partial \delta S}{\partial \theta}.\label{defkphi}
\end{eqnarray}
The transverse vorticity Eqs.~(\ref{dwthetadr}-\ref{dwphidr}) are rewritten in the equatorial plane using the definition (\ref{defktheta}-\ref{defkphi}) of $\delta k_\theta,\delta k_\varphi$ and the entropy Eq.~(\ref{dSdr}):
\begin{eqnarray}
\left(\frac{\partial}{\partial r}-\frac{i\omega'}{v_r}\right)\delta k_\theta
=
im \delta\left(\frac{{\cal L}}{\rho v_r}\right),\\
\left(\frac{\partial}{\partial r}-\frac{i\omega'}{v_r}\right)\delta k_\varphi
=r^2\delta w_r\frac{\partial\Omega}{\partial r}
-\frac{\partial}{\partial\theta}  \delta\left(\frac{{\cal L}}{\rho v_r}\right).
\end{eqnarray}
$\delta k_\theta$ is conserved in the adiabatic limit even with rotation. The conservation of $\delta k_\varphi$ in the adiabatic limit is valid with rotation only for the mirror symmetric solution such that $\delta w_r=0$.
We note that the adiabatic invariant $\delta K$ used in Paper~I without rotation is related to $\delta k_\theta,\delta k_\varphi$ as follows:
\begin{eqnarray}
\delta K&\equiv&r \left(\nabla\times\delta {\bf k}\right)_r,\\
&=&\frac{1}{\sin\theta} \frac{\partial}{\partial\theta} (\sin \theta \delta k_\varphi) -\frac{im}{\sin\theta} \delta k_\theta.
\end{eqnarray}
$\delta K$ is thus not conserved in an accretion flow with rotation, except for the mirror symmetric solution. \\
We define $\delta f$ as the perturbation  of the Bernoulli constant, incorporating the contribution of the azimuthal velocity $v_\varphi=L/(r\sin\theta)$. $\delta h$ is the perturbation of the radial mass flux and the entropy perturbation $\delta S$ is expressed with $\delta\rho$ and $\delta c^2$:
\begin{eqnarray}
\delta f &\equiv& v_r\delta v_r + \frac{L}{ r\sin\theta}\delta v_\varphi+
\frac{1}{\gamma-1}\delta c^2 \ , \label{defdf0}\\
\delta h&\equiv&\frac{\delta v_r}{v_r}+\frac{\delta \rho}{\rho},\label{defdg0}\\
\delta S&=&\frac{1}{\gamma-1}\frac{\delta c^2}{c^2}-\frac{\delta \rho}{\rho}.\label{defdS}
\end{eqnarray}
The relation between $\delta f$ and $\delta v_\varphi$ in the equatorial plane is deduced from the Euler equation in the direction $\varphi$ and using Eq.~(\ref{defktheta}):
\begin{eqnarray}
im \delta f
=
 i\omega r\delta v_\varphi 
+ \delta k_\theta.\label{dfk}
\end{eqnarray}
From Eqs.~(\ref{defdf0}), (\ref{defdg0}), (\ref{defdS}) and (\ref{dfk}) we express $\delta v_r$ and $\delta\rho$ with ($\delta h,\delta v_\varphi,\delta k_\theta,\delta S$)
in the equatorial plane :
\begin{eqnarray}
\frac{\delta v_r}{v_r}&=&\frac{1}{ 1-\M^2}\left(
\delta h+\delta S-\frac{\delta k_\theta}{ imc^2}-\frac{i\omega'}{c^2}\frac{r\delta v_\varphi}{im}
\right),\label{dvr_vr}\\
\frac{\delta\rho}{\rho}&=&\delta h-\frac{\delta v_r}{v_r}.\label{drho_rho}
\end{eqnarray}
The perturbed mass conservation is written as follows:
\begin{eqnarray}
v_r\frac{\partial \delta h}{ \partial r} = i\omega'\frac{\delta \rho}{\rho} - \frac{\delta A}{r^2}\label{mass_cons},
\end{eqnarray}
where $\delta A/r^2$ is the divergence of the ortho-radial velocity perturbation $(0,\delta v_\theta,\delta v_\varphi)$:
\begin{eqnarray}
\delta A&\equiv &\frac{r}{\sin\theta}\left\lbrack\frac{\partial}{\partial\theta}(\sin\theta\delta v_\theta)+
im  \delta v_\varphi\right\rbrack
\label{def_dA}.
\end{eqnarray}
In the equatorial plane, we use $\delta w_r=0$ to eliminate $\delta v_\theta$ from the definition (\ref{def_dA}) of $\delta A$:
\begin{eqnarray}
\delta v_\theta=\frac{1}{ im}\frac{\partial}{\partial\theta}(\sin\theta\delta v_\varphi),\\
\delta A= \frac{r}{im}\left\lbrack\frac{1}{\sin\theta}\frac{\partial}{\partial\theta}\left(\sin\theta\frac{\partial}{\partial\theta}(\sin\theta\delta v_\varphi)\right)
-\frac{m^2}{\sin^2\theta}  (\sin\theta\delta v_\varphi)\right\rbrack.\label{Laplacian}
\end{eqnarray}
We recognise in Eq.~(\ref{Laplacian}) the angular part of the Laplacian operator applied to $(\sin\theta\delta v_\varphi)$. Using its eigenvalue $-\ell(\ell+1)/r^2$ we obtain:
\begin{eqnarray}
\frac{r\delta v_\varphi}{im}&=&-\frac{\delta A}{\ell(\ell+1)}.\label{dAvphi}
\end{eqnarray}
Using Eq.~(\ref{dAvphi}), Eq.~(\ref{mass_cons}) is transformed into
\begin{eqnarray}
\frac{\partial \delta h}{ \partial r} &=& \frac{i\omega'}{v_r}\frac{\delta\rho}{\rho} + \frac{\ell(\ell+1)}{r^2v_r} \frac{r\delta v_\varphi}{im}.
\end{eqnarray}
The differential equation satisfied by $r\delta v_\varphi$ in the equatorial plane is deduced from the definition of the poloidal vorticity $\delta w_\theta$, the definition of $k_\theta$ in Eq.~(\ref{defktheta}) and Eq.~(\ref{dvr_vr}):
\begin{eqnarray}
\frac{\partial}{\partial r}\frac{r\delta v_\varphi}{ im}& = &
\delta v_r
-\frac{\delta k_\theta}{ imv_r}
+\frac{c^2}{v_r}\frac{\delta S}{\gamma}.\label{dvphik}
\end{eqnarray}
The differential system is expressed in Eqs.~(\ref{dhsh0}-\ref{dSsh0}) with ($\delta h,\delta v_\varphi,\delta k_\theta,\delta S$) using Eqs.~(\ref{dvr_vr}-\ref{drho_rho}).
The equations defining $\delta c^2$, $\delta p$, $\delta{\cal L}$ are repeated here for completeness:
\begin{eqnarray}
\frac{\delta c^2}{c^2}&=&(\gamma-1)\left(\frac{\delta \rho}{\rho}+\delta S\right),\label{dc2}\\
\frac{\delta p}{ \gamma p}&=&\frac{\delta \rho}{\rho}+\frac{\gamma-1}{\gamma}\delta S,\label{dP}\\
\delta \left( \frac{{\cal L}}{\rho v_r} \right) &=& \nabla S \frac{c^2}{\gamma} \left\lbrack (\beta - 1) \frac{\delta \rho}{\rho} + \alpha \frac{\delta c^2}{c^2} - \frac{\delta v_r}{v_r} \right\rbrack\ , \\
\delta \left( \frac{{\cal L}}{p v_r} \right) &=& \frac{\gamma}{ c^2}  \delta \left( \frac{\cal{L}}{\rho v_r} \right) - \frac{\delta c^2}{c^2} \nabla S\ .
\end{eqnarray}
The pressure perturbation can also be deduced from Eqs.~(\ref{defdf0}), (\ref{dfk}) and (\ref{dvphik}):
\begin{eqnarray}
\frac{\delta p}{ \gamma p}
&=&\frac{1}{ c^2}\left(
i\omega' 
-v_r\frac{\partial}{\partial r}
 \right) \frac{r\delta v_\varphi}{im}.\label{dpdvphi}
 \end{eqnarray} 
The boundary conditions required to numerically solve the linearized hydrodynamic equations are the same as in F07, modified by rotation as in YF08. \\
The condition at the inner boundary is $\delta v_{\rm ns}=0$.\\
As in WFT23 we introduce the effective potential $\Phi_{L^2}$ incorporating the centrifugal effect, and define the reference frequency $\omega_\Phi $ as follows:
\begin{eqnarray}
\Phi_{L^2}&\equiv&\Phi_0+\frac{L^2}{ 2r^2},\\
\frac{\partial \Phi_{L^2}}{\partial r}&=&\frac{\partial \Phi_0}{\partial r}-\frac{L^2}{ r^3},\\
\frac{\omega_\Phi r_{\rm sh}}{|v_{\rm sh}|}
&\equiv&
\frac{v_1}{\v2}
\frac{ \frac{r_{\rm sh}}{ v_1^2}\frac{\partial \Phi_{L^2}}{ \partial r}
-2\frac{\v2}{ v_1}
}{
1-\frac{\v2}{ v_{1}}
}
.
\end{eqnarray}
The boundary conditions satisfied by $r\delta v_\varphi,\delta h,\delta S$ at the shock displaced at $r_{\rm sh}+\Delta\zeta$ are the same as Eqs.~(A37-A40) in WFT23, repeated here for completeness:
\begin{eqnarray}
\frac{(r\delta v_\varphi)_{\rm sh}}{ im} =
\left(v_{1}-\v2\right)\Delta \zeta   ,\label{rdvphish}\\
\delta h_{\rm sh} =   -i\omega'
\left(1-\frac{\v2}{ v_{1}}\right)
\frac{\Delta \zeta}{ v_{\rm sh}}   \ ,\label{dhsh}\\
\delta S_{\rm sh}  = -\Delta \zeta 
\left\lbrack\nabla S\right\rbrack^{\rm sh}_{1}
+\gamma 
\frac{v_1}{c_{\rm sh}^2}
(i\omega' +\omega_\Phi)
\Delta\zeta
\left(1-\frac{\v2}{ v_{1}}\right)^2
\label{dSsh}
\ .
\end{eqnarray} 
Noting that $\delta f$ is not used in WFT23 and our definition (\ref{defdf0}) differs from Eq.~(15) in \cite{Buellet2023}, we calculate the value of $\delta f_{\rm sh}$ using the conservation of the energy flux in the frame of the shock at $r=r_{\rm sh}+\Delta\zeta$, defining the radial velocity of the shock as $\Delta v\equiv -i\omega\Delta \zeta$ and using Eq.~(A.36) in WFT23:
\begin{eqnarray}
\left\lbrack{\frac{1}{2}}(v_r-\Delta v+\delta v_r)^2+{\frac{1}{2}}\left(\frac{L}{ r}+\delta v_\varphi\right)^2+\frac{(c+\delta c)^2}{\gamma-1}
\right\rbrack_1^{\rm sh}=0,\\
\frac{\partial}{\partial r}\left(\frac{v_r^2}{2}+\frac{c^2}{\gamma-1}+\Phi_{L^2}\right)=\frac{{\cal L}}{\rho v_r}
\end{eqnarray} 
Linearizing, with $\Delta v=-i\omega\Delta\zeta$,
\begin{eqnarray}
\delta f_{\rm sh}=-(v_{\rm sh}-v_1)i\omega\Delta\zeta - \left\lbrack\frac{{\cal L}}{\rho v_r}\right\rbrack_1^{\rm sh}\Delta \zeta
.\label{rdfsh}
\end{eqnarray} 
The boundary condition satisfied by $\delta k_\theta$ at the shock is deduced from Eqs.~(\ref{dfk}), (\ref{rdvphish}) and (\ref{rdfsh}):
\begin{eqnarray}
\left(\frac{\delta k_\theta}{ im}\right)_{\rm sh}&=&\delta f_{\rm sh}-\frac{i\omega}{im}(r\delta v_\varphi)_{\rm sh},\\
&=&-\left\lbrack\frac{{\cal L}}{\rho v_r}\right\rbrack_1^{\rm sh}\Delta \zeta.\label{kthetash}
\end{eqnarray} 
The vorticity $\delta w_\theta$ produced at the shock is deduced from Eqs.~(\ref{defktheta}), (\ref{dSsh}) and (\ref{kthetash}), neglecting neutrino emission ahead of the shock:
\begin{eqnarray}
\frac{\delta w_{\rm sh}}{ im} &=&
\frac{1}{ r_{\rm sh} v_{\rm sh}}\left(\frac{\delta k_\theta}{ im}-\frac{c^2}{\gamma}\delta S\right)_{\rm sh},\\
&\sim&
-\frac{\Delta \zeta}{ r_{\rm sh}}\frac{v_1}{\v2}
\left(1-\frac{\v2}{ v_{1}}\right)^2
(i\omega'+\omega_\Phi)
 \label{vorticity_sh}.
 \end{eqnarray}

\section{Adiabatic model of equatorial accretion of a rotating gas}

\subsection{Second order differential system as a forced oscillator}
\label{app_forced_oscillator}

The expression for $\delta Y_{\rm sh}$ is deduced from Eqs.~(\ref{drvphish0}) and (\ref{def_Y}): 
\begin{eqnarray}
\frac{\delta Y_{\rm sh}}{ v_1}&=&\left(1-\frac{\v2}{ v_{1}}\right)\Delta\zeta.\label{Yshzeta}
\end{eqnarray} 
The expression for $\delta h_{\rm sh}$ and $\delta S_{\rm sh}$ are rewritten from Eqs.~(\ref{dhsh0}) and (\ref{dSsh0}) using Eq.~(\ref{Yshzeta}):
\begin{eqnarray}
\delta h_{\rm sh}&=&-i\omega'_{\rm sh}\frac{\delta Y_{\rm sh}}{ v_1v_{\rm sh}},\label{dhshY}\\
\frac{\delta S_{\rm sh}}{\gamma} &=& 
(i\omega'+\omega_\Phi)\left(1-\frac{\v2}{ v_1}\right) \frac{\delta Y_{\rm sh}}{ c_{\rm sh}^2}
\ .\label{dSshY}
\end{eqnarray} 
Using Eq.~(\ref{ddLdr}) with $\delta k_\theta=0$ and Eqs.~(\ref{dhshY}-\ref{dSshY}):
\begin{eqnarray}
\left(\frac{\partial \delta Y}{\partial X}\right)_{\rm sh}
=
a_1\frac{\delta Y_{\rm sh}}{ \v2^2}(i\omega'_{\rm sh}+\omega_\Phi)\left(1-\frac{\v2}{ v_1}\right)
,\label{dYsh_Ysh}\\
a_1\equiv
1+(\gamma-1)\M_{\rm sh}^2-\frac{i\omega'_{\rm sh}}{ i\omega'_{\rm sh}+\omega_\Phi}\frac{\frac{\v2}{ v_1}}{ 1-\frac{\v2}{ v_1}}.
\label{defa1}
\end{eqnarray} 
The inner boundary in the adiabatic approximation is deduced from Eq.~(\ref{inner_BC}) using Eq.~(\ref{ddLdr}):
\begin{eqnarray}
\left(\frac{\partial \delta Y}{ \partial X}\right)_{\rm ns}-\frac{i\omega'_{\rm ns}}{  c_{\rm ns}^2} {\delta Y_{\rm ns}}=
 \frac{1-\M^2_{\rm ns}}{\M^2_{\rm ns}}\frac{\delta S_{\rm sh}}{\gamma}{\rm e}^{\int_{\rm sh}^{\rm ns}\frac{i\omega'}{v_r}\frac{{\rm d}r}{1-\M^2}}
.\label{lowerBCY}
\end{eqnarray}

\subsection{Equation defining the eigenfrequencies in the adiabatic model}
\label{integral_dispersion}

The derivation of the integral equation defining the eigenfrequencies in the adiabatic model including rotation follows the same steps as in Appendix~D in Paper~I, where the following typo should be corrected: $\delta {\cal F}_{\rm sh}$ should be replaced by $\delta S_{\rm sh}/(\gamma\M_{\rm sh}^2)$ in Eqs.~(D10) to (D23). We extend this derivation to a rotating progenitor, incorporating the centrifugal correction in the effective potential and replacing $\omega$ with $\omega'(r)$. $\delta f/\omega$. We note $\delta {\cal F}\equiv {\cal F}\delta Y_{\rm sh}$.  $Y_0$ is defined as the solution of the homogeneous equation satisfying the inner boundary condition of pure acoustic waves (i.e. without entropy and vorticity perturbations), and $Y_-$ another homogeneous solution such that their Wronskien is noted $W$: 
\begin{eqnarray}
\left(\frac{\partial Y_0}{\partial X}\right)_{\rm ns}=\frac{i\omega'_{\rm ns}}{  c_{\rm ns}^2} Y_0^{\rm ns},\label{lowerbc0}\\
W\equiv Y_0\frac{\partial Y_-}{\partial X}-Y_-\frac{\partial Y_0}{\partial X}.\label{Wronskien}
\end{eqnarray}
The general solution $\delta Y$ of the differential equation (\ref{forced_oscillator_adiab}) and its derivative includes two integration constants $d_i$ and $d_0$ as follows:
\begin{eqnarray}
\delta Y=Y_-\left(d_-+\int_{\rm ns} \frac{Y_0}{ W} \delta{\cal F}{\rm d}X\right)
-Y_0\left(d_0+\int_{\rm sh} \frac{Y_-}{ W} \delta{\cal F}{\rm d}X\right),\label{Ygen}\\
\frac{\partial \delta Y}{\partial X}=\frac{\partial Y_-}{\partial X}\left(d_-+\int_{\rm ns} \frac{Y_0}{ W} \delta{\cal F}{\rm d}X\right)
\nonumber\\
-\frac{\partial Y_0}{\partial X}\left(d_0+\int_{\rm sh} \frac{Y_-}{ W} \delta{\cal F}{\rm d}X\right).\label{dYgen}
\end{eqnarray}
Using Eq.~(\ref{lowerbc0}) the lower boundary condition (\ref{lowerBCY}) translates into
\begin{eqnarray}
d_-\left\lbrack \left(\frac{\partial Y_-}{\partial X}\right)_{\rm ns} - \frac{i\omega'_{\rm ns}}{  c_{\rm ns}^2} Y_-^{\rm ns}\right\rbrack
=
\frac{\delta S_{\rm sh}}{\gamma}\frac{1-\M^2_{\rm ns}}{\M^2_{\rm ns}}{\rm e}^{\int_{\rm sh}^{\rm ns} \frac{i\omega'}{v_r}\frac{{\rm d}r}{1-\M^2}}.
\end{eqnarray}
We note using Eq.~(\ref{Wronskien}) and (\ref{lowerbc0}) that
\begin{eqnarray}
 \left(\frac{\partial Y_-}{\partial X}\right)_{\rm ns} - \frac{i\omega'_{\rm ns}}{  c_{\rm ns}^2} Y_-^{\rm ns}
  &=&\frac{W}{ Y_0^{\rm ns}}.
\end{eqnarray}
Thus
\begin{eqnarray}
Wd_-
=
Y_0^{\rm ns}
\frac{\delta S_{\rm sh}}{\gamma}\frac{1-\M^2_{\rm ns}}{\M^2_{\rm ns}}{\rm e}^{\int_{\rm sh}^{\rm ns}\frac{i\omega' }{v_r^2}{\rm d}X}.
\end{eqnarray}
Eliminating $d_0$ between Eqs.~(\ref{Ygen}) and (\ref{dYgen}) at the shock and using Eq.~(\ref{Wronskien}):
\begin{eqnarray}
Y_0^{\rm sh}\left(\frac{\partial \delta Y}{\partial X}\right)_{\rm sh}-\left(\frac{\partial Y_0}{\partial X}\right)_{\rm sh}\delta Y_{\rm sh}
=
Wd_-+\int_{\rm ns}^{\rm sh} Y_0 \delta{\cal F}{\rm d}X.
\end{eqnarray}
The eigenfrequencies are thus defined by
\begin{eqnarray}
Y_0^{\rm sh}\left(\frac{\partial \delta Y}{\partial X}\right)_{\rm sh}-\left(\frac{\partial Y_0}{\partial X}\right)_{\rm sh}\delta Y_{\rm sh}
=\nonumber\\
Y_0^{\rm ns}
\frac{\delta S_{\rm sh}}{\gamma}\frac{1-\M^2_{\rm ns}}{\M^2_{\rm ns}}{\rm e}^{\int_{\rm sh}^{\rm ns}\frac{i\omega' }{v_r^2}{\rm d}X}
+\int_{\rm ns}^{\rm sh} Y_0 \delta {\cal F}{\rm d}X.\label{interm}
\end{eqnarray}
Replacing the coupling term $\delta {\cal F}\equiv {\cal F}\delta Y_{\rm sh}$ by its expression (\ref{def_calF}), and using Eqs.~(\ref{dSshY}-\ref{dYsh_Ysh}), Eq.~(\ref{interm}) takes the following form
\begin{eqnarray}
a_1 Y_0^{\rm sh}
+a_2r_{\rm sh}\left(\frac{\partial Y_0}{\partial r}\right)_{\rm sh}
-(1-\M^2_{\rm ns})\frac{\M_{\rm sh}^2}{\M^2_{\rm ns}}{\rm e}^{\int_{\rm sh}^{\rm ns}  \frac{i\omega'}{v_r}\frac{{\rm d}r}{1-\M^2}}Y_0^{\rm ns}
=\nonumber\\
\int_{\rm ns}^{\rm sh} Y_0 
{\rm e}^{\int_{\rm sh} {i\omega'}\frac{{\rm d}X}{ c^2}}
\frac{\partial}{\partial r}
\left(
\frac{\M_{\rm sh}^2}{\M^2}
{\rm e}^{\int_{\rm sh} \frac{i\omega'}{v_r}{\rm d}r}
\right){\rm d}r,\label{dispersion0}
\end{eqnarray}
with $a_2$ defined by:
\begin{eqnarray}
a_2\equiv
-
\frac{\frac{\v2}{ r_{\rm sh}}}{ i\omega'_{\rm sh}+\omega_\Phi}
\frac{
1-\M_{\rm sh}^2
}{
1-\frac{\v2}{ v_{1}}
}
\label{defa2}
.
\end{eqnarray}
After one integration by parts, Eq.~(\ref{dispersion0}) becomes:
\begin{eqnarray}
(a_1-1)Y_0^{\rm sh}
+a_2r_{\rm sh}  \left(\frac{\partial Y_0}{\partial r}\right)_{\rm sh}
+\M_{\rm sh}^2{\rm e}^{\int_{\rm sh}^{\rm ns}\frac{i\omega'}{v_r}\frac{{\rm d}r}{1-\M^2}}Y_0^{\rm ns}
\nonumber\\
=-\int_{\rm ns}^{\rm sh} 
\frac{\partial}{\partial r}\left(
Y_0 
{\rm e}^{\int_{\rm sh} \frac{i\omega'\M^2}{1-\M^2}\frac{{\rm d}r}{v_r}}
\right)
\frac{\M_{\rm sh}^2}{\M^2}
{\rm e}^{\int_{\rm sh} \frac{i\omega'}{v_r}{\rm d}r}
{\rm d}r,\label{single_disp_app}\\
=
-\M_{\rm sh}^2
\left(
\int_{\rm ns}^{\rm sh} 
\frac{1-\M^2}{\M^2}\frac{\partial Y_0}{\partial r}
{\rm e}^{\int_{\rm sh} \frac{i\omega' }{1-\M^2}\frac{{\rm d}r}{v_r}}
{\rm d}r
\right.\nonumber\\
\left.
+
\left\lbrack
Y_0
{\rm e}^{\int_{\rm sh} \frac{i\omega' }{1-\M^2}\frac{{\rm d}r}{v_r}}
\right\rbrack_{\rm ns}^{\rm sh}
\right).
\label{single_disp_app2}
\end{eqnarray}
We note that without rotation, Eq.~(\ref{single_disp_app}) is the same as obtained in Paper~I (Eqs.~63-65, correcting a sign typo in the expression of $a_1-1$ in Eq.~64). \\
The more compact formulation in Eqs.~(\ref{single_disp}-\ref{def_b2}) results from developing and partially integrating the integral term, as shown in Eq.~(\ref{single_disp_app2}), and defining $b_2\equiv a_2/(1-\M_{\rm sh}^2)$ and $b_1\equiv a_1-1+\M_{\rm sh}^2$.

\end{appendix}

\end{document}